\documentclass[useAMS,usenatbib]{mn2e}

\usepackage{amssymb}
\usepackage{times}
\usepackage{graphicx}
\usepackage{epstopdf}
\usepackage{subfigure}
\usepackage[T1]{fontenc}
\usepackage{aecompl} 
\usepackage{hyperref}

\newcommand{\beq}{\begin{equation}}
\newcommand{\eeq}{\end{equation}}

\def\gs{\mathrel{\lower0.6ex\hbox{$\buildrel {\textstyle >}\over{\scriptstyle \sim}$}}}
\def\ls{\mathrel{\lower0.6ex\hbox{$\buildrel {\textstyle <}\over{\scriptstyle \sim}$}}}
\newcommand{\simgt}{\lower.5ex\hbox{$\; \buildrel > \over \sim \;$}}
\newcommand{\simlt}{\lower.5ex\hbox{$\; \buildrel < \over \sim \;$}}

\newcommand{\aap}{A\&A}
\newcommand{\apj}{ApJ}
\newcommand{\apjl}{ApJ}
\newcommand{\apjs}{ApJS}
\newcommand{\aj}{AJ}

\newcommand{\pasj}{PASJ}
\newcommand{\prd}{Phys. Rev. D}
\newcommand{\mnras}{MNRAS}

\newcommand{\ssr}{Space Science Reviews}

\begin{document}

\title[The $c$-$M$ relation of lensing clusters]{The mass--concentration relation in lensing clusters: the role of statistical biases and selection effects}
\author[
M. Sereno et al.
]{
Mauro Sereno$^{1,2}$\thanks{E-mail: mauro.sereno@unibo.it (MS)}, Carlo Giocoli$^{1,2,3}$, Stefano Ettori$^{2,3}$, Lauro Moscardini$^{1,2,3}$
\\
$^1$Dipartimento di Fisica e Astronomia, Alma Mater Studiorum -- Universit\`a di Bologna, Viale Berti Pichat 6/2, 40127 Bologna, Italia\\
$^2$INAF, Osservatorio Astronomico di Bologna, via Ranzani 1, 40127 Bologna, Italia\\
$^3$INFN, Sezione di Bologna, viale Berti Pichat 6/2, 40127 Bologna, Italia\\
}


\maketitle

\begin{abstract}
The relation between mass and concentration of galaxy clusters traces their formation and evolution. Massive lensing clusters were observed to be over-concentrated and following a steep scaling in tension with predictions from the concordance $\Lambda$CDM paradigm. We critically revise the relation in the CLASH, the SGAS, the LOCUSS, and the high-redshift samples of weak lensing clusters. Measurements of mass and concentration are anti-correlated, which can bias the observed relation towards steeper values. We corrected for this bias and compared the measured relation to theoretical predictions accounting for halo triaxiality, adiabatic contraction of the halo, presence of a dominant BCG and, mostly, selection effects in the observed sample. The normalisation, the slope and the scatter of the expected relation are strongly sample-dependent. For the considered samples, the predicted slope is much steeper than that of the underlying relation characterising dark-matter only clusters. We found that the correction for statistical and selection biases in observed relations mostly solve the tension with the $\Lambda$CDM model.
\end{abstract}

\begin{keywords}
galaxies: clusters: general -- dark matter -- gravitational lensing: weak -- gravitational lensing: strong
\end{keywords}

\section{Introduction}

The relation between mass and concentration of cluster of galaxies, $c$-$M$, is an important probe of the formation and evolution of matter haloes in the framework of the highly successful hierarchical cold dark matter model with a cosmological constant ($\Lambda$CDM). The concentration relates the density in the inner regions to the outer parts and it is connected to the mass and redshift of the halo \citep{bul+al01}. $N$-body simulations and theoretical models based on the mass accretion history show that concentrations are higher for lower mass halos and are smaller at early times \citep{bul+al01,duf+al08,zha+al09,gio+al12c}. A flattening of the $c$-$M$ relation occurs at higher masses and redshifts, but the extent of this feature is still debated \citep{kly+al11,pra+al11,lud+al12,lud+al14,me+ra13,du+ma14}.

Oddly, we observe a significant number of over-concentrated clusters \citep{bro+al08,ogu+al09,ume+al11b} and a steeper than predicted $c$-$M$ relation \citep{co+na07,fed12}. Inconsistencies among observations and predictions are shared by lensing and X-ray samples of clusters \citep{co+na07}. In the following, we mainly focus on lensing clusters.

Orientation and shape biases partially explain the over-concentration problem \citep{ogu+al05,ser+al10a,se+um11,ras+al12}. Efficient lenses are likely elongated towards the observer and their lensing strength is boosted \citep{hen+al07,og+bl09,men+al11}. Neglecting halo triaxiality can then lead to over-estimates up to a factor of two in concentration \citep{cor+al09}. The opposite takes place for lenses elongated in the plane of the sky. Corrections for shape and orientation require deep multi-wavelength observations \citep{ser+al12a,ser+al13,lim+al13}, which are very expensive to perform on large samples of clusters.

When stacking techniques are employed, the over-concentration problem is significantly reduced \citep{joh+al07,man+al08,cov+al14}. The stacked profile of samples of lensing clusters with steep $c$-$M$ relations is usually in line with theoretical predictions \citep{ogu+al12,se+co13,oka+al13}. Furthermore, the $c$-$M$ relation of stacked clusters of smaller mass is remarkably flat \citep{joh+al07,man+al08,cov+al14}. 

These results are still not conclusive. On the positive side, stacking techniques significantly increase the signal to noise ratio. They can probe the low mass regime and are less affected by projection effects. On the other hand, stacking brings a number of systematics mostly due to off-centring effects and averaging over a wide range of cluster properties which might affect the estimate of the concentration \citep{og+ta11,cov+al14}. Furthermore, cluster stacks may deviate from spherical symmetry. Weak lensing masses of stacked clusters selected by optical richness may be over-estimated by $\ga 5$ per cent even if clusters can be uniquely associated with haloes \citep{die+al14}.

Other sources of concern are related to centre offsets, selection effects, and shape noise from intrinsic ellipticities of background galaxies \citep{du+fa14}. The $c$-$M$ relation is usually determined in samples which are neither statistical nor complete and may constitute a biased population \citep{se+zi12}. Selection effects play a major role and can steepen the relation \citep{gio+al14}. Based on a suite of $N$-body/hydrodynamical simulations, \citet{men+al14} showed that the concentrations of CLASH clusters \citep[Cluster Lensing And Supernova survey with Hubble,][]{pos+al12} measured in \citet{mer+al14} are in line with theoretical predictions after accounting for projection and selection effects.

In this paper, we investigate whether the tension between observed $c$-$M$ relations in samples of weak lensing clusters and the $\Lambda$CDM paradigm can be reconciled with a proper treatment of the data and a better understanding of the selection effects. On the observational side, we take a critical approach at biases that affect the derivation of the $c$-$M$ relation. The mass and the concentration of a cluster are usually determined from the same data-set and through a single fitting procedure. The steep slope of the $c$-$M$ relation might be due to the strong anti-correlation between the measurements of halo mass and concentration \citep{aug+al13,du+ma14,du+fa14}.

On the theoretical side, we try to perform a consistent apples-to-apples comparison between the observed $c$-$M$ relation of lensing clusters and the theoretical predictions. The observed relation suffers from projection effects. Triaxiality and substructures can bias and scatter the observed lensing masses and concentrations, which differ from the intrinsic ones \citep{hen+al07,gio+al12b,se+et14_comalit_I}. Furthermore, processes due to baryons and their interplay with the total matter distribution affect the $c$-$M$ relation, which differ from expectations based on purely dark matter haloes \citep{fed12}. Finally, selection effects may preferentially include clusters with over-dense cores which can steepen the $c$-$M$ relation \citep{gio+al14,men+al14}. We explore these effects and try to correct for them when comparing observations to predictions with a semi-analytical approach based on the publicly available software MOKA \citep{gio+al12a}.

The paper is structured as follows. In Sec.~\ref{sec_samp}, we introduce the lensing samples and present the basic procedure used to recover masses and concentrations. In Sec.~\ref{sec_fit}, we determine the observed $c$-$M$ relations accounting for parameter anti-correlation in the framework of a Bayesian approach. Stacking techniques are employed in Sec.~\ref{sec_stac} to highlight the mean properties of the samples. Theoretical predictions for the $c$-$M$ relations tuned to the selection properties of the considered lensing samples and accounting for projection effects and baryonic physics are derived and compared to observations in Sec.~\ref{sec_theo}. Discussion of further sources of disagreement is contained in Sec.~\ref{sec_disc}. Section~\ref{sec_conc} is devoted to some final considerations.

Throughout the paper, we assume a flat $\Lambda$CDM cosmology with density parameters $\Omega_\mathrm{M}=0.3$, $\Omega_{\Lambda}=0.7$ and Hubble constant $H_0=100h~\mathrm{km~s}^{-1}\mathrm{Mpc}^{-1}$. When necessary, we fixed $h=0.7$ and $\sigma_8=0.81$ for the amplitude of the matter power spectrum.

\section{Cluster samples}
\label{sec_samp}

\begin{table*}
\caption{Masses, $M_{200}$, and concentrations, $c_{200}$, of the CLASH sample. Estimates in cols. 2--7 (8--13) were obtained under the assumption of uniform (log-uniform) priors. We report the cluster names (col.~1), the masses, $M_{200}$ (cols. 2 or 8), the concentrations $M_{200}$ (cols. 3 or 9), the correlations between $M_{200}$ and $c_{200}$, $\delta_{Mc}$ (cols. 4 or 10), the decimal logarithms of the mass, $\log M_{200}$ (cols. 5 and 11), the decimal logarithms of the concentration, $\log c_{200}$ (cols. 6 and 12), and the correlations between $\log M_{200}$ and $\log c_{200}$, $\delta_{\log Mc}$ (cols. 7 and 13). Reported values are the bi-weight estimators of the marginalised distributions. Masses are in units of $10^{14}M_\odot/h$.}
\label{tab_clash}
\resizebox{\hsize}{!} {
\begin{tabular}[c]{l  r@{$\,\pm\,$}l r@{$\,\pm\,$}l c r@{$\,\pm\,$}l r@{$\,\pm\,$}l c r@{$\,\pm\,$}l r@{$\,\pm\,$}l c r@{$\,\pm\,$}l r@{$\,\pm\,$}l c}
\hline
        \noalign{\smallskip}
         &  \multicolumn{10}{c}{Uniform priors}  &  \multicolumn{10}{c}{Log-uniform priors} \\
	Name	&  \multicolumn{2}{c}{$M_{200}$} & \multicolumn{2}{c}{$c_{200}$} & $\delta_{Mc}$ &  \multicolumn{2}{c}{$\log M_{200}$} & \multicolumn{2}{c}{$\log c_{200}$} & $\delta_{\log Mc}$ &  \multicolumn{2}{c}{$M_{200}$} & \multicolumn{2}{c}{$c_{200}$} & $\delta_{Mc}$ &  \multicolumn{2}{c}{$\log M_{200}$} & \multicolumn{2}{c}{$\log c_{200}$} & $\delta_{\log Mc}$  \\
		&  \multicolumn{2}{c}{$[10^{14}M_\odot/h]$} &  \multicolumn{2}{c}{} &  &  \multicolumn{2}{c}{} &  \multicolumn{2}{c}{} &  &  \multicolumn{2}{c}{$[10^{14}M_\odot/h]$} &  \multicolumn{2}{c}{}  &   &  \multicolumn{2}{c}{}  &  \multicolumn{2}{c}{}  &    \\
        \noalign{\smallskip}
        \hline
 ABELL 383 & 5.5 & 1.5 & 7.0 & 1.2 & -0.60 & 0.73 & 0.12 & 0.85 & 0.07 & -0.95 & 5.2 & 1.4 & 7.2 & 1.2 & -0.62 & 0.72 & 0.12 & 0.85 & 0.07 & -0.95 \\
ABELL 209 & 16.2 & 2.6 & 3.0 & 0.3 & -0.35 & 1.21 & 0.07 & 0.48 & 0.04 & -0.93 & 16.0 & 2.6 & 3.0 & 0.3 & -0.35 & 1.20 & 0.07 & 0.48 & 0.04 & -0.93 \\
ABELL 2261 & 13.8 & 1.7 & 4.8 & 0.5 & -0.50 & 1.14 & 0.05 & 0.68 & 0.04 & -0.85 & 13.7 & 1.7 & 4.8 & 0.5 & -0.50 & 1.14 & 0.05 & 0.68 & 0.04 & -0.85 \\
RX J2129.3+0005 & 4.3 & 0.8 & 6.7 & 0.9 & -1.08 & 0.63 & 0.08 & 0.83 & 0.06 & -0.90 & 4.2 & 0.8 & 6.8 & 0.9 & -1.10 & 0.62 & 0.08 & 0.83 & 0.06 & -0.90 \\
ABELL 611 & 10.2 & 2.5 & 4.5 & 0.8 & -0.35 & 1.01 & 0.11 & 0.65 & 0.08 & -0.94 & 10.5 & 2.5 & 4.4 & 2.1 & -0.19 & 1.02 & 0.11 & 0.65 & 0.10 & -0.26 \\
MS 2137.3-2353 & 7.1 & 1.7 & 5.2 & 0.9 & -0.53 & 0.85 & 0.10 & 0.72 & 0.08 & -0.93 & 7.0 & 1.7 & 5.3 & 0.9 & -0.54 & 0.84 & 0.10 & 0.72 & 0.08 & -0.93 \\
RXC J2248.7-4431 & 8.8 & 3.0 & 7.1 & 2.6 & -0.28 & 0.94 & 0.15 & 0.85 & 0.15 & -0.94 & 8.8 & 3.1 & 7.0 & 2.6 & -0.26 & 0.94 & 0.16 & 0.85 & 0.16 & -0.94 \\
MACS J1115.8+0129 & 10.5 & 2.7 & 4.1 & 0.8 & -0.34 & 1.02 & 0.11 & 0.61 & 0.08 & -0.94 & 10.5 & 3.1 & 4.1 & 0.9 & -0.29 & 1.02 & 0.13 & 0.61 & 0.09 & -0.95 \\
MACS J1931.8-2635 & 8.6 & 5.4 & 5.9 & 3.3 & -0.14 & 0.91 & 0.27 & 0.77 & 0.22 & -0.98 & 8.2 & 4.5 & 6.1 & 2.9 & -0.17 & 0.90 & 0.24 & 0.78 & 0.20 & -0.97 \\
RX J1532.9+3021 & 3.3 & 1.1 & 7.5 & 2.3 & -0.72 & 0.51 & 0.14 & 0.88 & 0.13 & -0.86 & 3.2 & 1.0 & 7.5 & 2.4 & -0.74 & 0.50 & 0.14 & 0.88 & 0.14 & -0.85 \\
MACS J1720.3+3536 & 7.1 & 1.8 & 6.1 & 1.5 & -0.46 & 0.85 & 0.12 & 0.78 & 0.11 & -0.93 & 6.9 & 1.8 & 6.2 & 1.5 & -0.46 & 0.84 & 0.12 & 0.79 & 0.10 & -0.92 \\
MACS J0429.6-0253 & 5.5 & 1.7 & 5.8 & 1.4 & -0.46 & 0.74 & 0.13 & 0.76 & 0.11 & -0.95 & 5.4 & 1.6 & 5.8 & 1.4 & -0.54 & 0.73 & 0.13 & 0.76 & 0.10 & -0.94 \\
MACS J1206.2-0847 & 9.3 & 2.3 & 6.3 & 1.8 & -0.36 & 0.97 & 0.11 & 0.80 & 0.12 & -0.92 & 9.6 & 2.3 & 6.1 & 1.6 & -0.37 & 0.98 & 0.10 & 0.78 & 0.11 & -0.91 \\
MACS J0329.6-0211 & 6.1 & 1.1 & 9.0 & 2.3 & -0.77 & 0.79 & 0.08 & 0.95 & 0.11 & -0.86 & 6.1 & 1.0 & 8.9 & 2.2 & -0.77 & 0.79 & 0.07 & 0.95 & 0.11 & -0.85 \\
RX J1347.5-1145 & 19.5 & 5.0 & 4.5 & 1.2 & -0.17 & 1.29 & 0.11 & 0.66 & 0.12 & -0.93 & 19.3 & 4.9 & 4.5 & 1.2 & -0.16 & 1.28 & 0.11 & 0.66 & 0.12 & -0.93 \\
MACS J0744.9+3927 & 9.0 & 2.3 & 7.1 & 2.6 & -0.34 & 0.95 & 0.11 & 0.85 & 0.15 & -0.89 & 9.1 & 2.3 & 6.9 & 2.3 & -0.34 & 0.95 & 0.11 & 0.84 & 0.15 & -0.90 \\
MACS J0416.1-2403 & 6.0 & 1.0 & 8.0 & 1.7 & -0.79 & 0.78 & 0.08 & 0.90 & 0.09 & -0.85 & 6.0 & 1.0 & 7.9 & 1.7 & -0.80 & 0.78 & 0.07 & 0.90 & 0.09 & -0.84 \\
MACS J1149.5+2223 & 18.6 & 3.3 & 2.7 & 0.4 & -0.26 & 1.27 & 0.08 & 0.44 & 0.07 & -0.89 & 18.6 & 3.4 & 2.7 & 0.4 & -0.26 & 1.27 & 0.08 & 0.44 & 0.07 & -0.90 \\
MACS J0717.5+3745 & 19.9 & 2.9 & 5.2 & 1.2 & -0.24 & 1.30 & 0.06 & 0.71 & 0.10 & -0.71 & 20.0 & 2.9 & 5.0 & 1.2 & -0.23 & 1.30 & 0.06 & 0.70 & 0.10 & -0.69 \\
MACS J0647.7+7015 & 8.0 & 2.1 & 7.3 & 2.6 & -0.38 & 0.90 & 0.12 & 0.87 & 0.15 & -0.90 & 8.1 & 2.1 & 7.1 & 2.4 & -0.39 & 0.90 & 0.11 & 0.85 & 0.14 & -0.89 \\
\hline
\end{tabular}
}
\end{table*}

\begin{table*}
\caption{Same as Table~\ref{tab_clash}, but for the SGAS sample.}
\label{tab_sgas}
\resizebox{\hsize}{!} {
\begin{tabular}[c]{l  r@{$\,\pm\,$}l r@{$\,\pm\,$}l c r@{$\,\pm\,$}l r@{$\,\pm\,$}l c r@{$\,\pm\,$}l r@{$\,\pm\,$}l c r@{$\,\pm\,$}l r@{$\,\pm\,$}l c}
\hline
        \noalign{\smallskip}
         &  \multicolumn{10}{c}{Uniform priors}  &  \multicolumn{10}{c}{Log-uniform priors} \\
	Name	&  \multicolumn{2}{c}{$M_{200}$} & \multicolumn{2}{c}{$c_{200}$} & $\delta_{Mc}$ &  \multicolumn{2}{c}{$\log M_{200}$} & \multicolumn{2}{c}{$\log c_{200}$} & $\delta_{\log Mc}$ &  \multicolumn{2}{c}{$M_{200}$} & \multicolumn{2}{c}{$c_{200}$} & $\delta_{Mc}$ &  \multicolumn{2}{c}{$\log M_{200}$} & \multicolumn{2}{c}{$\log c_{200}$} & $\delta_{\log Mc}$  \\
		&  \multicolumn{2}{c}{$[10^{14}M_\odot/h]$} &  \multicolumn{2}{c}{} &  &  \multicolumn{2}{c}{} &  \multicolumn{2}{c}{} &  &  \multicolumn{2}{c}{$[10^{14}M_\odot/h]$} &  \multicolumn{2}{c}{}  &   &  \multicolumn{2}{c}{}  &  \multicolumn{2}{c}{}  &    \\
        \noalign{\smallskip}
        \hline
SDSS J0851+3331 & 5.0 & 1.4 & 8.6 & 2.6 & -0.57 & 0.70 & 0.12 & 0.94 & 0.13 & -0.92 & 5.0 & 1.4 & 8.6 & 2.6 & -0.60 & 0.69 & 0.12 & 0.93 & 0.13 & -0.91 \\
SDSS J0915+3826 & 1.3 & 0.5 & 13.2 & 3.8 & -1.57 & 0.12 & 0.15 & 1.11 & 0.13 & -0.92 & 1.3 & 0.5 & 13.4 & 3.7 & -1.68 & 0.11 & 0.15 & 1.12 & 0.13 & -0.90 \\
SDSS J0957+0509 & 1.4 & 0.8 & 6.9 & 2.6 & -0.95 & 0.12 & 0.26 & 0.84 & 0.16 & -0.98 & 1.1 & 0.7 & 7.9 & 3.2 & -1.17 & 0.03 & 0.26 & 0.90 & 0.17 & -0.98 \\
SDSS J1004+4112 & 1.8 & 1.3 & 8.5 & 4.4 & -0.60 & 0.25 & 0.28 & 0.91 & 0.21 & -0.98 & 1.6 & 1.2 & 9.1 & 4.6 & -0.59 & 0.21 & 0.28 & 0.93 & 0.22 & -0.98 \\
SDSS J1029+2623 & 2.0 & 0.5 & 9.2 & 4.3 & -1.23 & 0.29 & 0.12 & 0.95 & 0.20 & -0.70 & 2.0 & 0.5 & 8.1 & 4.0 & -1.17 & 0.29 & 0.12 & 0.90 & 0.21 & -0.63 \\
SDSS J1038+4849 & 1.2 & 0.3 & 14.8 & 3.5 & -1.97 & 0.07 & 0.12 & 1.17 & 0.11 & -0.84 & 1.2 & 0.3 & 14.5 & 3.6 & -1.95 & 0.07 & 0.13 & 1.16 & 0.11 & -0.85 \\
SDSS J1050+0017 & 5.4 & 1.3 & 10.1 & 4.3 & -0.52 & 0.73 & 0.11 & 0.99 & 0.19 & -0.71 & 5.5 & 1.4 & 8.7 & 4.0 & -0.46 & 0.74 & 0.11 & 0.93 & 0.19 & -0.71 \\
RCS2 J1055+5547 & 4.1 & 1.0 & 6.4 & 1.1 & -0.91 & 0.61 & 0.10 & 0.80 & 0.08 & -0.92 & 4.1 & 1.0 & 6.4 & 1.1 & -0.92 & 0.61 & 0.10 & 0.81 & 0.08 & -0.92 \\
SDSS J1110+6459 & 3.1 & 1.3 & 12.2 & 4.9 & -0.39 & 0.49 & 0.19 & 1.08 & 0.20 & -0.68 & 2.9 & 1.3 & 11.3 & 4.9 & -0.39 & 0.47 & 0.20 & 1.04 & 0.21 & -0.56 \\
SDSS J1115+5319 & 8.7 & 2.5 & 4.6 & 1.7 & -0.26 & 0.93 & 0.13 & 0.66 & 0.17 & -0.71 & 8.6 & 2.5 & 4.3 & 1.6 & -0.24 & 0.93 & 0.13 & 0.63 & 0.17 & -0.65 \\
SDSS J1138+2754 & 8.9 & 0.5 & 3.6 & 0.3 & -0.84 & 0.95 & 0.02 & 0.55 & 0.03 & -0.42 & 8.8 & 0.5 & 3.6 & 0.3 & -0.85 & 0.95 & 0.03 & 0.55 & 0.03 & -0.43 \\
SDSS J1152+3313 & 0.9 & 0.4 & 12.5 & 4.0 & -1.78 & -0.04 & 0.19 & 1.09 & 0.14 & -0.95 & 0.9 & 0.4 & 12.9 & 4.0 & -1.70 & -0.06 & 0.19 & 1.10 & 0.14 & -0.95 \\
SDSS J1152+0930 & 4.9 & 1.8 & 3.1 & 1.6 & -0.36 & 0.69 & 0.16 & 0.49 & 0.24 & -0.60 & 4.6 & 1.8 & 2.3 & 1.8 & -0.21 & 0.66 & 0.18 & 0.36 & 0.42 & -0.11 \\
SDSS J1209+2640 & 5.4 & 1.5 & 6.7 & 1.3 & -0.60 & 0.73 & 0.12 & 0.83 & 0.09 & -0.94 & 5.3 & 1.4 & 6.8 & 1.3 & -0.63 & 0.72 & 0.12 & 0.83 & 0.08 & -0.93 \\
SDSS J1226+2149 & 7.0 & 2.4 & 4.9 & 1.5 & -0.32 & 0.84 & 0.15 & 0.69 & 0.13 & -0.89 & 6.9 & 2.3 & 4.9 & 1.4 & -0.34 & 0.84 & 0.15 & 0.69 & 0.12 & -0.89 \\
SDSS J1226+2152 & 1.3 & 1.5 & 6.2 & 5.8 & -0.30 & 0.11 & 0.48 & 0.72 & 0.48 & -0.67 & 0.5 & 0.7 & 0.7 & 1.5 & -0.25 & -0.36 & 0.62 & -0.12 & 0.96 & -0.25 \\
ABELL 1703 & 9.5 & 1.4 & 5.6 & 0.8 & -0.62 & 0.98 & 0.06 & 0.75 & 0.06 & -0.85 & 9.4 & 1.4 & 5.6 & 0.8 & -0.61 & 0.97 & 0.06 & 0.75 & 0.06 & -0.85 \\
SDSS J1315+5439 & 3.7 & 1.0 & 12.3 & 4.5 & -0.49 & 0.57 & 0.11 & 1.08 & 0.18 & -0.46 & 3.7 & 1.0 & 11.0 & 4.6 & -0.45 & 0.56 & 0.12 & 1.03 & 0.20 & -0.41 \\
GHO 132029+3155 & 2.7 & 0.4 & 12.3 & 3.2 & -1.66 & 0.42 & 0.07 & 1.08 & 0.11 & -0.70 & 2.7 & 0.4 & 12.0 & 3.1 & -1.62 & 0.42 & 0.07 & 1.07 & 0.11 & -0.70 \\
SDSS J1329+2243 & 4.7 & 0.9 & 5.0 & 0.9 & -0.87 & 0.67 & 0.09 & 0.70 & 0.07 & -0.84 & 4.6 & 1.0 & 5.0 & 0.9 & -0.86 & 0.66 & 0.09 & 0.70 & 0.08 & -0.85 \\
SDSS J1343+4155 & 3.2 & 1.1 & 4.5 & 1.3 & -0.64 & 0.49 & 0.16 & 0.65 & 0.13 & -0.78 & 3.0 & 1.1 & 4.5 & 1.4 & -0.59 & 0.47 & 0.17 & 0.65 & 0.14 & -0.37 \\
SDSS J1420+3955 & 6.2 & 1.9 & 4.3 & 1.1 & -0.41 & 0.79 & 0.13 & 0.63 & 0.11 & -0.90 & 6.1 & 1.8 & 4.2 & 1.1 & -0.44 & 0.78 & 0.13 & 0.63 & 0.11 & -0.89 \\
SDSS J1446+3032 & 4.4 & 1.1 & 10.0 & 3.9 & -0.60 & 0.64 & 0.11 & 0.99 & 0.17 & -0.79 & 4.5 & 1.2 & 9.2 & 3.6 & -0.61 & 0.65 & 0.11 & 0.96 & 0.17 & -0.76 \\
SDSS J1456+5702 & 3.0 & 0.5 & 15.6 & 2.8 & -1.34 & 0.47 & 0.08 & 1.19 & 0.08 & -0.78 & 3.0 & 0.6 & 15.5 & 2.8 & -1.35 & 0.47 & 0.08 & 1.19 & 0.08 & -0.79 \\
SDSS J1531+3414 & 4.5 & 1.1 & 6.8 & 1.2 & -0.80 & 0.65 & 0.11 & 0.83 & 0.08 & -0.92 & 4.4 & 1.0 & 6.9 & 1.2 & -0.83 & 0.64 & 0.11 & 0.84 & 0.07 & -0.92 \\
SDSS J1621+0607 & 5.0 & 1.6 & 4.4 & 1.0 & -0.53 & 0.70 & 0.14 & 0.64 & 0.10 & -0.90 & 4.9 & 1.6 & 4.4 & 1.0 & -0.51 & 0.69 & 0.14 & 0.64 & 0.10 & -0.90 \\
SDSS J1632+3500 & 3.5 & 1.0 & 10.9 & 4.4 & -0.56 & 0.54 & 0.13 & 1.02 & 0.18 & -0.63 & 3.5 & 1.1 & 9.6 & 4.3 & -0.52 & 0.54 & 0.13 & 0.97 & 0.19 & -0.55 \\
SDSS J2111-0114 & 4.0 & 1.9 & 5.1 & 3.8 & -0.33 & 0.59 & 0.21 & 0.70 & 0.31 & -0.68 & 3.7 & 2.0 & 3.4 & 3.4 & -0.18 & 0.56 & 0.24 & 0.54 & 0.50 & 0.15 \\
\hline
\end{tabular}
}
\end{table*}

\begin{table*}
\caption{Same as Table~\ref{tab_clash}, but for the high-$z$ sample.}
\label{tab_highz}
\resizebox{\hsize}{!} {
\begin{tabular}[c]{l  r@{$\,\pm\,$}l r@{$\,\pm\,$}l c r@{$\,\pm\,$}l r@{$\,\pm\,$}l c r@{$\,\pm\,$}l r@{$\,\pm\,$}l c r@{$\,\pm\,$}l r@{$\,\pm\,$}l c}
\hline
        \noalign{\smallskip}
         &  \multicolumn{10}{c}{Uniform priors}  &  \multicolumn{10}{c}{Log-uniform priors} \\
	Name	&  \multicolumn{2}{c}{$M_{200}$} & \multicolumn{2}{c}{$c_{200}$} & $\delta_{Mc}$ &  \multicolumn{2}{c}{$\log M_{200}$} & \multicolumn{2}{c}{$\log c_{200}$} & $\delta_{\log Mc}$ &  \multicolumn{2}{c}{$M_{200}$} & \multicolumn{2}{c}{$c_{200}$} & $\delta_{Mc}$ &  \multicolumn{2}{c}{$\log M_{200}$} & \multicolumn{2}{c}{$\log c_{200}$} & $\delta_{\log Mc}$  \\
		&  \multicolumn{2}{c}{$[10^{14}M_\odot/h]$} &  \multicolumn{2}{c}{} &  &  \multicolumn{2}{c}{} &  \multicolumn{2}{c}{} &  &  \multicolumn{2}{c}{$[10^{14}M_\odot/h]$} &  \multicolumn{2}{c}{}  &   &  \multicolumn{2}{c}{}  &  \multicolumn{2}{c}{}  &    \\
        \noalign{\smallskip}
        \hline
XMMXCS J2215-1738 & 1.9 & 1.1 & 8.7 & 5.9 & -0.28 & 0.30 & 0.26 & 0.90 & 0.38 & -0.71 & 2.9 & 3.3 & 2.1 & 3.8 & -0.11 & 0.49 & 0.46 & 0.23 & 0.82 & -0.60 \\
XMMU J2205-0159 & 1.4 & 0.9 & 4.7 & 3.3 & -0.60 & 0.12 & 0.28 & 0.67 & 0.30 & -0.76 & 1.3 & 0.9 & 3.8 & 3.1 & -0.38 & 0.12 & 0.32 & 0.58 & 0.38 & -0.53 \\
XMMU J1229+0151 & 23.9 & 16.0 & 0.5 & 0.6 & -0.04 & 1.32 & 0.41 & -0.34 & 0.56 & -0.88 & 13.2 & 12.5 & 0.9 & 1.0 & -0.05 & 1.06 & 0.40 & -0.16 & 0.64 & -0.87 \\
WARPS J1415+3612 & 2.3 & 1.6 & 4.0 & 2.1 & -0.35 & 0.36 & 0.29 & 0.59 & 0.23 & -0.98 & 2.4 & 1.8 & 3.9 & 2.1 & -0.33 & 0.37 & 0.31 & 0.59 & 0.24 & -0.99 \\
ISCS J1432+3332 & 5.1 & 3.7 & 2.0 & 1.5 & -0.06 & 0.73 & 0.36 & 0.30 & 0.39 & -0.91 & 5.6 & 5.0 & 1.7 & 1.4 & -0.10 & 0.75 & 0.34 & 0.23 & 0.43 & -0.83 \\
ISCS J1429+3437 & 27.3 & 19.7 & 0.4 & 0.5 & -0.03 & 1.36 & 0.42 & -0.39 & 0.56 & -0.82 & 14.3 & 17.7 & 0.5 & 0.8 & -0.03 & 1.07 & 0.52 & -0.32 & 0.79 & -0.83 \\
ISCS J1434+3427 & 1.6 & 1.3 & 5.9 & 6.1 & -0.24 & 0.23 & 0.36 & 0.66 & 0.54 & -0.74 & 2.3 & 3.4 & 0.6 & 1.2 & -0.06 & 0.37 & 0.63 & -0.18 & 0.89 & -0.45 \\
ISCS J1432+3436 & 2.4 & 1.9 & 5.3 & 5.5 & -0.25 & 0.39 & 0.32 & 0.65 & 0.46 & -0.81 & 7.8 & 11.7 & 0.5 & 0.9 & -0.03 & 0.85 & 0.59 & -0.25 & 0.91 & -0.81 \\
ISCS J1434+3519 & 7.8 & 11.1 & 0.7 & 0.9 & -0.03 & 0.88 & 0.55 & -0.17 & 0.67 & -0.80 & 4.7 & 6.9 & 0.3 & 0.5 & -0.05 & 0.61 & 0.66 & -0.50 & 0.74 & -0.37 \\
ISCS J1438+3414 & 1.8 & 1.4 & 6.9 & 5.9 & -0.25 & 0.28 & 0.34 & 0.75 & 0.47 & -0.66 & 1.3 & 1.5 & 1.5 & 2.7 & -0.12 & 0.09 & 0.60 & 0.08 & 0.88 & 0.06 \\
RCS 0220-0333 & 2.8 & 1.5 & 5.6 & 4.4 & -0.30 & 0.45 & 0.23 & 0.72 & 0.33 & -0.87 & 3.9 & 4.7 & 3.1 & 3.5 & -0.10 & 0.66 & 0.41 & 0.43 & 0.55 & -0.90 \\
RCS 0221-0321 & 1.1 & 0.4 & 11.6 & 5.2 & -1.04 & 0.03 & 0.18 & 1.05 & 0.22 & -0.56 & 1.0 & 0.5 & 9.4 & 5.6 & -0.74 & 0.01 & 0.21 & 0.95 & 0.31 & -0.38 \\
RCS 0337-2844 & 4.1 & 3.7 & 3.3 & 3.9 & -0.09 & 0.62 & 0.38 & 0.51 & 0.48 & -0.85 & 4.4 & 4.2 & 2.2 & 2.8 & -0.09 & 0.64 & 0.40 & 0.32 & 0.63 & -0.69 \\
RCS 0439-2904 & 2.1 & 0.9 & 7.3 & 4.9 & -0.35 & 0.32 & 0.20 & 0.83 & 0.30 & -0.83 & 2.1 & 1.0 & 6.1 & 4.1 & -0.57 & 0.33 & 0.19 & 0.77 & 0.29 & -0.75 \\
RCS 2156-0448 & 0.6 & 0.5 & 7.9 & 5.4 & -0.52 & -0.18 & 0.38 & 0.86 & 0.31 & -0.96 & 0.6 & 0.4 & 8.2 & 4.9 & -0.70 & -0.24 & 0.35 & 0.88 & 0.27 & -0.94 \\
RCS 1511+0903 & 1.1 & 0.5 & 11.4 & 5.5 & -0.81 & 0.04 & 0.19 & 1.05 & 0.25 & -0.49 & 1.0 & 0.6 & 6.8 & 6.0 & -0.39 & 0.01 & 0.30 & 0.77 & 0.52 & -0.22 \\
RCS 2345-3632 & 1.6 & 1.2 & 2.7 & 2.6 & -0.19 & 0.21 & 0.35 & 0.42 & 0.48 & -0.85 & 1.8 & 1.5 & 1.7 & 1.9 & -0.18 & 0.26 & 0.38 & 0.19 & 0.66 & -0.70 \\
RCS 2319+0038 & 1.7 & 0.6 & 11.5 & 4.8 & -0.97 & 0.24 & 0.16 & 1.04 & 0.20 & -0.93 & 1.8 & 0.7 & 10.6 & 4.7 & -0.62 & 0.25 & 0.17 & 1.01 & 0.21 & -0.94 \\
XLSS J0223-0436 & 17.8 & 15.1 & 0.8 & 0.9 & -0.04 & 1.18 & 0.41 & -0.08 & 0.53 & -0.90 & 40.2 & 26.3 & 0.1 & 0.2 & -0.02 & 1.59 & 0.40 & -0.76 & 0.69 & -0.85 \\
RDCS J0849+4452 & 2.4 & 1.1 & 2.8 & 1.0 & -0.70 & 0.38 & 0.19 & 0.45 & 0.15 & -0.94 & 2.3 & 1.0 & 2.9 & 1.0 & -0.74 & 0.36 & 0.19 & 0.46 & 0.15 & -0.93 \\
RDCS J0910+5422 & 2.3 & 0.9 & 6.6 & 3.9 & -0.64 & 0.37 & 0.16 & 0.81 & 0.25 & -0.87 & 2.4 & 0.9 & 5.8 & 3.4 & -0.72 & 0.38 & 0.16 & 0.76 & 0.25 & -0.85 \\
RDCS J1252-2927 & 4.4 & 1.2 & 4.6 & 1.7 & -0.60 & 0.64 & 0.12 & 0.66 & 0.17 & -0.90 & 4.5 & 1.1 & 4.4 & 1.5 & -0.65 & 0.65 & 0.11 & 0.64 & 0.15 & -0.87 \\
XMM J2235-2557 & 8.9 & 8.2 & 2.1 & 1.4 & -0.07 & 1.01 & 0.34 & 0.33 & 0.32 & -0.88 & 8.1 & 5.6 & 2.4 & 1.3 & -0.12 & 0.94 & 0.28 & 0.39 & 0.21 & -0.80 \\
CL J1226+3332 & 7.1 & 1.6 & 4.3 & 1.1 & -0.49 & 0.85 & 0.10 & 0.64 & 0.11 & -0.90 & 7.1 & 1.6 & 4.3 & 1.1 & -0.51 & 0.85 & 0.10 & 0.63 & 0.11 & -0.90 \\
MS 1054-0321 & 16.1 & 11.5 & 0.9 & 0.8 & -0.06 & 1.19 & 0.29 & -0.08 & 0.42 & -0.92 & 23.7 & 15.2 & 0.6 & 0.7 & -0.06 & 1.31 & 0.31 & -0.37 & 0.62 & -0.89 \\
CL J0152-1357 & 2.0 & 0.4 & 11.3 & 3.9 & -1.80 & 0.29 & 0.09 & 1.04 & 0.15 & -0.86 & 2.0 & 0.4 & 10.5 & 3.5 & -1.85 & 0.30 & 0.09 & 1.01 & 0.14 & -0.86 \\
RDCS J0848+4453 & 1.5 & 1.7 & 2.8 & 4.4 & -0.21 & 0.21 & 0.44 & 0.40 & 0.62 & -0.87 & 1.8 & 2.0 & 1.1 & 1.7 & -0.17 & 0.24 & 0.46 & -0.02 & 0.78 & -0.58 \\
XMMU J1230.3+1339 & 17.1 & 10.8 & 2.1 & 2.1 & -0.06 & 1.19 & 0.32 & 0.35 & 0.46 & -0.65 & 20.9 & 22.1 & 0.5 & 0.7 & -0.01 & 1.31 & 0.51 & -0.35 & 0.79 & -0.36 \\
ClG J1604+4304 & 2.8 & 1.7 & 9.4 & 5.8 & -0.23 & 0.45 & 0.27 & 0.94 & 0.33 & -0.43 & 1.9 & 2.2 & 1.9 & 3.7 & -0.05 & 0.24 & 0.68 & 0.14 & 0.92 & 0.03 \\
RX J1716.6+6708 & 3.5 & 2.4 & 5.0 & 4.7 & -0.16 & 0.57 & 0.29 & 0.65 & 0.43 & -0.87 & 5.5 & 6.3 & 2.2 & 2.9 & -0.09 & 0.75 & 0.39 & 0.25 & 0.72 & -0.81 \\
ClG 1137.5+6625 & 6.5 & 3.2 & 3.6 & 1.9 & -0.18 & 0.81 & 0.22 & 0.56 & 0.23 & -0.95 & 7.1 & 3.6 & 3.2 & 1.7 & -0.10 & 0.85 & 0.24 & 0.51 & 0.23 & -0.90 \\
\hline
\end{tabular}
}
\end{table*}

\begin{table*}
\caption{Same as Table~\ref{tab_clash}, but for the  LOCUSS sample.}
\label{tab_locuss}
\resizebox{\hsize}{!} {
\begin{tabular}[c]{l  r@{$\,\pm\,$}l r@{$\,\pm\,$}l c r@{$\,\pm\,$}l r@{$\,\pm\,$}l c r@{$\,\pm\,$}l r@{$\,\pm\,$}l c r@{$\,\pm\,$}l r@{$\,\pm\,$}l c}
\hline
        \noalign{\smallskip}
         &  \multicolumn{10}{c}{Uniform priors}  &  \multicolumn{10}{c}{Log-uniform priors} \\
	Name	&  \multicolumn{2}{c}{$M_{200}$} & \multicolumn{2}{c}{$c_{200}$} & $\delta_{Mc}$ &  \multicolumn{2}{c}{$\log M_{200}$} & \multicolumn{2}{c}{$\log c_{200}$} & $\delta_{\log Mc}$ &  \multicolumn{2}{c}{$M_{200}$} & \multicolumn{2}{c}{$c_{200}$} & $\delta_{Mc}$ &  \multicolumn{2}{c}{$\log M_{200}$} & \multicolumn{2}{c}{$\log c_{200}$} & $\delta_{\log Mc}$  \\
		&  \multicolumn{2}{c}{$[10^{14}M_\odot/h]$} &  \multicolumn{2}{c}{} &  &  \multicolumn{2}{c}{} &  \multicolumn{2}{c}{} &  &  \multicolumn{2}{c}{$[10^{14}M_\odot/h]$} &  \multicolumn{2}{c}{}  &   &  \multicolumn{2}{c}{}  &  \multicolumn{2}{c}{}  &    \\
        \noalign{\smallskip}
        \hline
  ABELL 68 & 3.1 & 1.2 & 4.9 & 3.8 & -0.39 & 0.49 & 0.16 & 0.69 & 0.33 & -0.73 & 3.2 & 1.3 & 3.5 & 2.9 & -0.37 & 0.50 & 0.18 & 0.55 & 0.37 & -0.59 \\
ABELL 115 & 4.5 & 2.8 & 2.7 & 2.7 & -0.12 & 0.65 & 0.29 & 0.43 & 0.48 & -0.83 & 4.8 & 3.0 & 1.5 & 1.7 & -0.15 & 0.67 & 0.27 & 0.14 & 0.62 & -0.53 \\
ZwCl 0104.4+0048 & 1.4 & 0.4 & 7.9 & 4.7 & -1.46 & 0.14 & 0.11 & 0.88 & 0.25 & -0.56 & 1.4 & 0.4 & 5.7 & 3.8 & -1.26 & 0.15 & 0.12 & 0.75 & 0.27 & -0.48 \\
ABELL 209 & 10.1 & 1.9 & 2.1 & 0.5 & -0.35 & 1.00 & 0.08 & 0.33 & 0.11 & -0.71 & 10.1 & 1.9 & 2.1 & 0.5 & -0.35 & 1.00 & 0.08 & 0.31 & 0.11 & -0.71 \\
RX J0142.0+2131 & 4.2 & 0.9 & 5.4 & 1.6 & -0.74 & 0.62 & 0.10 & 0.73 & 0.13 & -0.73 & 4.2 & 1.0 & 5.1 & 1.6 & -0.71 & 0.62 & 0.10 & 0.71 & 0.14 & -0.71 \\
ABELL 267 & 3.0 & 0.7 & 5.1 & 1.8 & -0.95 & 0.47 & 0.11 & 0.71 & 0.15 & -0.73 & 3.0 & 0.7 & 4.8 & 1.6 & -0.90 & 0.47 & 0.11 & 0.68 & 0.15 & -0.68 \\
ABELL 291 & 5.4 & 1.6 & 1.8 & 0.9 & -0.34 & 0.73 & 0.13 & 0.25 & 0.24 & -0.68 & 5.7 & 1.8 & 1.4 & 0.9 & -0.30 & 0.76 & 0.14 & 0.14 & 0.29 & -0.65 \\
ABELL 383 & 2.9 & 0.7 & 7.7 & 3.7 & -1.01 & 0.46 & 0.11 & 0.88 & 0.20 & -0.84 & 3.0 & 0.7 & 6.9 & 3.4 & -0.97 & 0.47 & 0.11 & 0.84 & 0.21 & -0.83 \\
ABELL 521 & 4.8 & 1.2 & 1.6 & 0.4 & -0.55 & 0.68 & 0.11 & 0.21 & 0.12 & -0.81 & 4.8 & 1.2 & 1.6 & 0.4 & -0.60 & 0.68 & 0.11 & 0.20 & 0.12 & -0.80 \\
ABELL 586 & 15.8 & 9.3 & 1.0 & 0.4 & -0.10 & 1.17 & 0.26 & -0.03 & 0.18 & -0.99 & 12.7 & 5.8 & 1.0 & 0.3 & -0.15 & 1.09 & 0.20 & 0.02 & 0.14 & -0.99 \\
ZwCl 0740.4+1740 & 4.9 & 3.1 & 1.8 & 1.5 & -0.05 & 0.70 & 0.35 & 0.25 & 0.45 & -0.91 & 5.6 & 4.7 & 1.4 & 1.3 & -0.12 & 0.75 & 0.31 & 0.11 & 0.51 & -0.81 \\
ZwCl 0823.2+0425 & 5.5 & 1.5 & 2.6 & 1.2 & -0.43 & 0.74 & 0.12 & 0.41 & 0.21 & -0.70 & 5.7 & 1.5 & 2.2 & 1.1 & -0.39 & 0.75 & 0.12 & 0.34 & 0.23 & -0.62 \\
ZwCl 0839.9+2937 & 2.0 & 0.6 & 8.2 & 4.4 & -1.10 & 0.29 & 0.12 & 0.90 & 0.23 & -0.69 & 2.1 & 0.6 & 6.5 & 3.7 & -1.02 & 0.31 & 0.13 & 0.80 & 0.24 & -0.67 \\
ABELL 611 & 5.2 & 1.2 & 3.7 & 1.4 & -0.51 & 0.72 & 0.10 & 0.56 & 0.16 & -0.65 & 5.3 & 1.1 & 3.4 & 1.3 & -0.51 & 0.72 & 0.09 & 0.53 & 0.17 & -0.59 \\
ABELL 689 & 1.1 & 0.6 & 1.5 & 1.6 & -0.70 & 0.04 & 0.22 & 0.19 & 0.54 & -0.39 & 0.9 & 0.6 & 0.2 & 0.2 & -0.04 & -0.08 & 0.33 & -0.74 & 0.66 & 0.14 \\
ABELL 697 & 9.2 & 1.7 & 2.5 & 0.7 & -0.32 & 0.96 & 0.08 & 0.39 & 0.13 & -0.54 & 9.2 & 1.7 & 2.3 & 0.7 & -0.29 & 0.96 & 0.08 & 0.36 & 0.14 & -0.50 \\
ABELL 750 & 7.9 & 3.0 & 2.5 & 1.4 & -0.20 & 0.90 & 0.16 & 0.39 & 0.24 & -0.86 & 8.7 & 3.5 & 2.1 & 1.3 & -0.18 & 0.93 & 0.17 & 0.31 & 0.27 & -0.83 \\
ABELL 963 & 5.2 & 1.1 & 2.0 & 0.7 & -0.59 & 0.72 & 0.09 & 0.29 & 0.15 & -0.66 & 5.3 & 1.1 & 1.8 & 0.7 & -0.53 & 0.72 & 0.09 & 0.25 & 0.16 & -0.67 \\
ABELL 1835 & 10.5 & 2.3 & 2.7 & 0.8 & -0.29 & 1.02 & 0.10 & 0.44 & 0.13 & -0.73 & 10.5 & 2.3 & 2.6 & 0.8 & -0.28 & 1.02 & 0.10 & 0.42 & 0.13 & -0.67 \\
ZwCl 1454.8+2233 & 2.6 & 1.4 & 3.6 & 2.8 & -0.43 & 0.40 & 0.23 & 0.56 & 0.35 & -0.70 & 2.9 & 1.8 & 2.1 & 2.3 & -0.28 & 0.45 & 0.27 & 0.30 & 0.60 & -0.39 \\
ABELL 2009 & 3.0 & 0.8 & 5.5 & 1.9 & -0.85 & 0.47 & 0.12 & 0.74 & 0.15 & -0.77 & 3.0 & 0.8 & 5.2 & 1.8 & -0.81 & 0.47 & 0.12 & 0.71 & 0.15 & -0.75 \\
ZwCl 1459.4+4240 & 2.6 & 0.8 & 8.3 & 3.9 & -0.66 & 0.42 & 0.14 & 0.91 & 0.20 & -0.60 & 2.7 & 0.9 & 6.9 & 3.6 & -0.59 & 0.42 & 0.14 & 0.83 & 0.23 & -0.49 \\
ABELL 2219 & 8.2 & 1.9 & 6.6 & 2.9 & -0.37 & 0.91 & 0.10 & 0.81 & 0.19 & -0.81 & 8.3 & 2.0 & 5.9 & 2.5 & -0.36 & 0.92 & 0.10 & 0.77 & 0.18 & -0.80 \\
RX J1720.1+2638 & 3.2 & 1.0 & 8.5 & 4.1 & -0.65 & 0.51 & 0.13 & 0.92 & 0.21 & -0.77 & 3.3 & 1.1 & 7.5 & 3.7 & -0.61 & 0.52 & 0.14 & 0.87 & 0.21 & -0.75 \\
ABELL 2261 & 12.4 & 0.8 & 4.1 & 0.6 & -0.75 & 1.09 & 0.03 & 0.61 & 0.06 & -0.58 & 12.4 & 0.7 & 4.0 & 0.6 & -0.73 & 1.09 & 0.03 & 0.60 & 0.06 & -0.55 \\
ABELL 2345 & 17.2 & 6.5 & 0.2 & 0.1 & -0.09 & 1.22 & 0.17 & -0.82 & 0.38 & -0.56 & 16.6 & 7.4 & 0.1 & 0.1 & -0.05 & 1.22 & 0.19 & -1.11 & 0.40 & -0.45 \\
RX J2129.6+0005 & 6.8 & 2.5 & 2.0 & 1.2 & -0.24 & 0.83 & 0.16 & 0.30 & 0.27 & -0.70 & 7.0 & 2.6 & 1.5 & 1.1 & -0.21 & 0.84 & 0.16 & 0.18 & 0.33 & -0.53 \\
ABELL 2390 & 6.8 & 1.5 & 5.3 & 1.3 & -0.48 & 0.83 & 0.10 & 0.72 & 0.11 & -0.76 & 6.7 & 1.5 & 5.2 & 1.3 & -0.47 & 0.83 & 0.10 & 0.71 & 0.11 & -0.75 \\
ABELL 2485 & 3.4 & 1.2 & 3.1 & 1.8 & -0.37 & 0.53 & 0.15 & 0.49 & 0.26 & -0.67 & 3.4 & 1.2 & 2.4 & 1.5 & -0.41 & 0.53 & 0.15 & 0.38 & 0.29 & -0.45 \\
ABELL 2631 & 4.5 & 0.8 & 6.6 & 2.5 & -0.87 & 0.65 & 0.08 & 0.82 & 0.16 & -0.72 & 4.6 & 0.8 & 5.9 & 2.2 & -0.82 & 0.66 & 0.08 & 0.77 & 0.16 & -0.72 \\
\hline
\end{tabular}
}
\end{table*}

We measured masses and concentrations for a number of lensing clusters with publicly available reduced shear profile, which we briefly introduce in this section. We refer to the original papers for a detailed presentation of the data reduction and analysis. For most of the clusters in the samples, masses and concentrations have been measured in the original papers. We re-determined them for a number of reasons. Firstly, we wanted to analyse each cluster with the same procedure and within the same $\Lambda$CDM reference model. Secondly, we needed the estimates of the correlation between measured mass and concentration, which were not provided in the original papers. Thirdly, we needed the mean value of the logarithm of mass and concentration, which are usually unpublished, rather than the logarithm of the mean value. Finally, some values of mass or concentration were missing.

Dark matter haloes can be conveniently described as Navarro-Frenk-White (NFW) density profiles \citep{nav+al97,ji+su02},
\begin{equation}
\label{nfw1}
	\rho_\mathrm{NFW}=\frac{\rho_\mathrm{s}}{(r/r_\mathrm{s})(1+r/r_\mathrm{s})^2},
\end{equation}
where $r_\mathrm{s}$ is the scale radius and $\rho_\mathrm{s}$ is four times the density at $r_\mathrm{s}$. The evolution of the concentration with time and mass exhibits the smallest deviations from universality whether halo masses are defined with respect to the critical density of the universe \citep{di+kr15}. The radius $r_{200}$ is defined such that the mean density contained within it is 200 times the critical density at the halo redshift. The corresponding concentration is $c_{200} \equiv r_{200}/ r_\mathrm{s}$. $M_{200}$ is the mass inside such sphere.

The observed shear profiles were fitted to spherical NFW models. As free parameters, we considered the mass $M_{200}$ and the concentration $c_{200}$. In terms of a $\chi^2$ function,
\beq
\label{eq_chi_WL}
\chi_\mathrm{WL}^2 ( M_{200},c_{200} )=\sum_i \left[ \frac{g_{+}(\theta_i)-g_{+}^\mathrm{NFW}(\theta_i; M_{200},c_{200})}{\delta_{+}(\theta_i)}\right]^2,
\eeq
where $g_{+}$  is the reduced tangential shear measured in circular annuli at angular position $\theta_i$ and $\delta_{+}$ is the corresponding observational uncertainty. Expressions for the  shear induced by a NFW halo can be found in \citet{wr+br00}. We considered the fitting radial range proposed in the original papers, which usually comprises all the measured shear data points. If suggested we removed the innermost radial bin in the reduced shear profile \citep{jee+al11,se+co13}.

The weak lensing observables in the cluster regime are nonlinearly related to the underlying lensing potential. Under the single source-plane approximation, nonlinear corrections to the source-averaged reduced shear profile have to be applied \citep[and references therein]{ume13}. In typical observations, the effect is usually negligible. 

Let us consider a massive cluster with $M_{200}\sim 10^{15}M_\odot/h$ and $c_{200}\sim 3$ at $z\sim0.3$. For a mean redshift of the background sources of $z_\mathrm{s}\sim 1$ and a factor $f_{W}=\langle W^2\rangle/\langle W\rangle^2$ of $\sim 1.05$, where $W$ is the relative lensing strength, the reduced shear is under-estimated by less than 1 (0.03) per cent at an angular radius $\theta \sim 1\arcmin$ ($15\arcmin$).

Similarly, the reduced shear should be corrected also for averaging in annular bins. As before, the effect is small. Considering the same lensing system as before and 10 bins equally spaced in logarithmic space, the reduced shear at the weighted averaged radius under-estimates the averaged reduced shear by less that 0.2 (0.8) per cent in the annulus $1.0\arcmin <\theta<1.3\arcmin$ ($13\arcmin <\theta<15\arcmin$).

These effects are much smaller than the observational uncertainties and the information needed to correct for them is not always available from the original papers where the shear measurements were performed. We did not correct for them.

When available, we also considered the strong lensing (SL) constraint on the Einstein radius through
\beq
\label{eq_chi_SL}
\chi_\mathrm{SL}^2 ( M_{200},c_{200} )=
\left[\frac{
\theta_\mathrm{E}-\theta_\mathrm{E}^\mathrm{NFW}(M_{200},c_{200})}{\delta \theta_\mathrm{E}}
\right]^2,
\eeq
where $\theta_\mathrm{E}$ is the measured effective angular Einstein radius. 

The combined $\chi^2$ is 
\beq
\label{eq_chi_GL}
\chi_\mathrm{GL}^2=\chi_\mathrm{SL}^2+\chi_\mathrm{WL}^2,
\eeq
and the likelihood is ${\cal L} \propto \exp \{-(\chi^2_\mathrm{WL}+\chi^2_\mathrm{SL})/2\}$. 

We considered two kinds of priors for $M_{200}$ and $c_{200}$. As a reference case, we considered uniform priors in the ranges $0.02 \le M_{200}/(10^{14}h^{-1}M_\odot) \le 100$ and $0.02 \le c_{200} \le 20$ (`uniform' priors). Throughout this section, for comparison with other works, we consider masses and concentrations obtained with these priors.

As an alternative, we considered priors uniform in logarithmically spaced intervals within the same bounds (`log-uniform' priors), as suitable for positive parameters. Virial masses determined with weak lensing are accurate whereas the estimates of the concentration are more uncertain. Log-uniform priors may avoid the bias of the concentration toward large values that can plague lensing analysis of low-quality data \citep{se+co13}.

\subsection{CLASH}

The Cluster Lensing And Supernova survey with Hubble \citep[CLASH,][]{pos+al12} has been providing very deep multi-wavelength observations of 25 massive clusters drawn largely from the Abell and the MAssive Cluster Survey \citep[MACS,][]{ebe+al10} cluster catalogs.

The X-ray selected clusters of the sample are fairly luminous with X-ray temperatures $T_\mathrm{X} \ge 5$~keV and show a smooth morphology in their X-ray surface brightness. The off-sets between the X-ray luminosity centroid and the brightest cluster galaxy (BCG) is  $ \ls 30$ kpc \citep{ume+al14}. By comparison with a sample of simulated halos which resemble the X-ray morphology, \citet{men+al14} showed that the X-ray selected CLASH clusters are mainly relaxed halos, but the sample also contains a significant fraction of un-relaxed systems.

\citet{ume+al14} presented the weak lensing analysis of a sub-sample of 16 X-ray regular and 4 high-magnification galaxy clusters in the redshift range $0.19 \ls z \ls 0.69$. Mass estimates were obtained with joint weak lensing shear plus magnification measurements based on ground-based wide-field Subaru data. 

\citet{mer+al14} combined weak lensing constraints from the Hubble Space Telescope (HST) and from Subaru with strong lensing constraints from HST. They measured mass and concentration for 19 CLASH clusters with regular X-ray morphology and found $c_\mathrm{200}\propto (1+z)^{-0.14\pm0.52} M_\mathrm{200}^{-0.32\pm0.18}$.

We re-analysed the shear profiles of the 20 clusters with ground-based measurements in \citet{ume+al14}. The strong lensing analysis of the CLASH clusters and the values of the mean distances from the cluster centre to the critical line, which we used as effective Einstein radii, can be found in \citet{zit+al14}. Results are summarised in Table~\ref{tab_clash}. We report the results obtained either assuming priors uniform in linear space or in logarithmic decades.

The choice of the priors for the analysis of high quality data has a minor impact. Mass and concentration estimates obtained assuming prior distributions which are either linearly uniform or uniform in logarithmic decades differ by less than 1 per cent, with a scatter smaller than 2 per cent. We also tested that the different maximum allowed parameter values, in particular the upper bound on the concentration, have a negligible effect on the estimates.

Together with the estimates of mass and concentration, we also reported in Table~\ref{tab_clash} the estimates of their logarithms, which we adopted for the fitting of the $c$-$M$ relation. In fact, for shallow or complex distributions $\log \langle M_{200}\rangle$ or $\log \langle c_{200}\rangle$ may be very different from $\langle \log M_{200} \rangle$ or $\langle \log c_{200} \rangle$, respectively. This is not the case of the CLASH sample. Posterior distributions are regularly shaped and with a pronounced peak and there is no meaningful difference between the logarithm of the mean and the mean of the logarithm.

Notwithstanding the differences in the data-sets, our mass determinations are consistent with \citet{ume+al14}, who considered log-uniform priors in the ranges $1 < M_\mathrm{200}/(10^{14}h^{-1}M_\odot) < 100$ and $0.1 < c_{200} < 10$. With respect to their results, our values of $M_{200}$ are slightly smaller by $\sim 12 \pm 15$ per cent. We computed typical deviations and scatters as the bi-weight estimators of the distribution of (un-weighted) relative mass differences. For the comparison, we uniformed the cosmological models as described in \citet{ser14_comalit_III}. The small discrepancy may be due to the different data-sets. \citet{ume+al14} used ground based shear and magnification measurements, whereas we used ground based shear measurements together with strong lensing constraints. 

Our mass determinations are in good agreement also with \citet{mer+al14}. The relative difference in $M_{200}$ is $\sim 0 \pm 35$ per cent. On the other hand, our determinations of $c_{200}$ are larger by $40 \pm 30$ per cent. The analysis in \citet{mer+al14} weighted more the strong lensing data and also considered the shear in the intermediate regime probed by HST. This can explain the differences in the estimated concentrations.

\subsection{SGAS}

The Sloan Giant Arcs Survey \citep[SGAS,][]{hen+al08} is a survey of strongly lensed giant arcs from the Sloan Digital Sky Survey \citep[SDSS,][]{sdss_york_et_al_00}. \citet{ogu+al12} presented a combined strong and weak lensing analysis for a subsample of 28 clusters in the redshift range $0.27 \la z \la 0.68$, based on follow-up imaging observations with Subaru/Suprime-cam. They found a steep relation $c_\mathrm{vir}\propto M_\mathrm{vir}^{-0.59\pm0.12}$. Based on the same mass determinations, \citet{fed12} found $c_\mathrm{200}\propto M_\mathrm{200}^{-0.67\pm0.13}$.

We re-analyzed the shear profiles of the 28 clusters published in \citet{ogu+al12} with the additional constraints on the effective Einstein radii \citep[ table 2]{ogu+al12}. Results are reported in Table~\ref{tab_sgas}. As for the CLASH sample, results are independent of priors. Masses (concentrations) obtained with uniformly logarithmic priors are smaller by just $1\pm2$  ($1\pm6$) per cent. $\log \langle M_{200}\rangle$  ($\log \langle c_{200}\rangle$) is not distinguishable from $\langle \log M_{200} \rangle$ ($\langle \log c_{200} \rangle$).

In order to derive masses and concentrations,  \citet{ogu+al12} adopted a $\chi^2$ function analog to ours, see Eqs.~(\ref{eq_chi_WL},~\ref{eq_chi_SL}, and~\ref{eq_chi_GL}). The main difference is that they employed a maximum-likelihood rather than a Bayesian approach. They performed the fitting in the parameter range $0.1 < M_\mathrm{vir}/(10^{14}h^{-1}M_\odot) <100 $ and $0.01 < c_\mathrm{vir} < 39.81$. 

Our results are highly consistent with the analysis in \citet{ogu+al12}. For the comparison, we translated the values quoted in \citet{ogu+al12}, which refers to the virial radius, to an over-density of $\Delta=200$ \citep{hu+kr03} and to our reference cosmological parameters \citep{ser14_comalit_III}. We found that our values of $M_{200}$ are smaller by $\sim 1\pm11$ per cent, whereas our concentrations $c_{200}$ are larger by $\sim 1\pm19$ per cent.

\subsection{High-$z$}

\citet{se+co13} collected from literature a heterogeneous sample of 31 high-redshift galaxy clusters at $0.8 \la z \la 1.5$ with measured shear profile (high-$z$ sample in the following). The clusters were mostly discovered within either X-ray or optical surveys. Furthermore, the shear measurements were performed by different groups.

Given the variety of finding techniques, we do not expect that all discoveries were affected by the same bias. Furthermore, biases due to orientation or projection of large-scale structure are strongly mitigated for a large sample. The clusters were not selected based on their lensing strength and they should not be severely affected by biases plaguing lensing-selected samples, such as the over-concentration problem and the orientation bias \citep{og+bl09}. Furthermore, the orientation bias is limited for X-ray- selected clusters \citep{men+al11}. These positive factors counterbalance the heterogeneous nature of the sample.

Results are reported in Table~\ref{tab_highz}. Masses are independent of priors whereas the effect on concentration is significant, as expected for data with a smaller signal to noise ratio. Masses obtained with log-uniform priors are larger by 3 $\pm$ 29 per cent. Concentrations are smaller by 28 $\pm$ 40 per cent. The differences between $\log \langle M_{200}\rangle$  ($\log \langle c_{200}\rangle$) and $\langle \log M_{200} \rangle$ ($\langle \log c_{200} \rangle$) are slightly larger than for the other samples, but still small. $\log \langle c_{200}\rangle$ is larger than $\langle \log c_{200} \rangle$ by 1$\pm$ 3 per cent.

The differences in the fitting procedure of mass and concentration with respect to \citet{se+co13} are minimal and results are fully consistent.  \citet{se+co13} found $c_\mathrm{200}\propto M_\mathrm{200}^{-0.83\pm0.39}$.

\subsection{LOCUSS}

The Local Cluster Substructure Survey (LOCUSS) team presented a Subaru weak lensing analysis of 30 X-ray luminous galaxy clusters at $0.15 \la z \la 0.30$ \citep{oka+al10}. Clusters were selected by the requirement to be bright enough in the X-ray band and lie above the ROSAT All Sky Survey \citep[RASS,][]{ebe+al00} flux limit, irrespective of their dynamical status. 

LOCUSS masses presented in \citet{oka+al10} are biased low due to contamination effects and systematics in shape measurements \citep{oka+al13}. The underestimate of $M_{200}$ ($c_{200}$) is of the order of 20 (20) per cent with a scatter of 14 (19) per cent \citep{oka+al13}. We cannot exclude mass dependent effects. The $c$-$M$ based on the LOCUSS sample is then significantly biased. Nevertheless, the masses quoted in \citet{oka+al10} are routinely used to study cluster properties \citep{planck_int_III,mar+al14}. We considered the LOCUSS sample to highlight the effects of hidden assumptions in linear regression. 

Among the full sample of 30 clusters, \citet{oka+al10} further selected a sub-sample of 19 morphologically regular clusters with spectroscopic redshift, a filter coverage in two bands and a smooth shear profile. We refer to this sub-sample as `LOCUSS regular'. For the LOCUSS regular sample, \citet{oka+al10} found $c_\mathrm{vir}\propto M_\mathrm{vir}^{-0.40\pm0.19}$.  

Results of our analysis for the full sample of 30 clusters are summarised in Table~\ref{tab_locuss}. Mass estimates are independent of priors. $M_{200}$'s obtained with uniformly logarithmic priors are larger by just 1 $\pm$ 3 per cent. The effect of priors is more pronounced on concentrations. $c_{200}$'s obtained with uniformly logarithmic priors are smaller by 12 $\pm$ 15 per cent. $\log \langle M_{200}\rangle$  ($\log \langle c_{200}\rangle$) is not distinguishable from $\langle \log M_{200} \rangle$ ($\langle \log c_{200} \rangle$).

\citet[ see their table 6]{oka+al10} quoted virial masses and concentrations for 26 clusters. We converted these values to an over-density radius $r_{200}$ as described for the SGAS sample. Results are fully consistent. Our values for $M_{200}$ are slightly smaller by $\sim 4\pm12$ ($\sim 8\pm20$) per cent.  Our $c_{200}$'s are slightly larger by $\sim 5\pm21$ per cent.

\section{Observed relations}
\label{sec_fit}

\begin{table*}
\caption{The observed $c$-$M$ relations. Col.~1: sample. Col.~2: type of regression. Cols.~3 and 4: type of prior on $M_{200}$ and $c_{200}$, respectively. `unif.': the prior is a uniform distribution; `log-unif.': the prior is a uniform distribution in logarithmic decades; `Student': the prior is a Student's $t$ distribution. Col.~5: type of prior on the slope $\beta$. Col. 6: assumption on the error matrix covariance. `$\delta_{Mc}\neq 0$ (=0)' means that we considered (neglected) the correlation between observed masses and concentrations. Col.~7: number of clusters in the sample, $N_\mathrm{cl}$. Col.~8: median redshift of the sample. Col.~9: pivot mass, $M_\mathrm{pivot}$, in units of $10^{14}M_\odot/h$, see Eq.~(\ref{eq_cM_1}). Cols. 10, 11, and 12: normalisation, slope, and intrinsic scatter of the $c$-$M$ relation, see Eq.~(\ref{eq_cM_1}). Quoted values are bi-weight estimators of the posterior probability distribution.}
\label{tab_par_obs}
\resizebox{\hsize}{!} {
\begin{tabular}[c]{l l l l l c	c c c r@{$\,\pm\,$}lr@{$\,\pm\,$}lr@{$\,\pm\,$}l}
        \hline
        \noalign{\smallskip}
	Sample		& Fit	&	$M$-prior	&	$c$-prior	&	$\beta$-prior	&	Covariance	&	$N_\mathrm{cl}$	&	$z$	&	$M_\mathrm{pivot}$	& \multicolumn{2}{c}{$\alpha$} & \multicolumn{2}{c}{$\beta$} & \multicolumn{2}{c}{$\sigma_{\log}$}  \\
	\multicolumn{8}{c}{}	&	$[10^{14}M_\odot/h]$	& \multicolumn{5}{c}{}  \\
         \hline   
	CLASH			&	fit-lin		&	unif.		&	unif.		&	Student	& $\delta_{Mc}\neq 0$	&	20	&	0.38	&	8.71	&	0.73	&	0.02	&	-0.46	&	0.11	&	0.08	&	0.02	\\
					&	fit-log	&	log-unif.	&	log-unif.	&	Student	& $\delta_{Mc}\neq 0$	&	"	&	"	&	8.50	&	0.73	&	0.02	&	-0.43	&	0.11	&	0.07	&	0.02	\\
					&	fit-bias	&	unif.		&	unif.		&	unif.		& $\delta_{Mc}= 0$		&	"	&	"	&	8.71	&	0.73	&	0.03	&	-0.59	&	0.16	&	0.05	&	0.02	\\
	\hline
	SGAS			&	fit-lin		&	unif.		&	unif.		&	Student	& $\delta_{Mc}\neq 0$	&	28	&	0.46	&	4.05	&	0.82	&	0.03	&	-0.40	&	0.14	&	0.12	&	0.02	\\
					&	fit-log	&	log-unif.	&	log-unif.	&	Student	& $\delta_{Mc}\neq 0$	&	"	&	"	&	3.89	&	0.83	&	0.03	&	-0.44	&	0.13	&	0.12	&	0.02	\\
					&	fit-bias	&	unif.		&	unif.		&	unif.		& $\delta_{Mc}= 0$		&	"	&	"	&	4.05	&	0.86	&	0.03	&	-0.73	&	0.15	&	0.07	&	0.03	\\
	\hline
	high-$z$			&	fit-lin		&	unif.		&	unif.		&	Student	& $\delta_{Mc}\neq 0$	&	31	&	1.07	&	2.42	&	0.69	&	0.06	&	-0.27	&	0.35	&	0.12	&	0.05	\\
					&	fit-log	&	log-unif.	&	log-unif.	&	Student	& $\delta_{Mc}\neq 0$	&	"	&	"	&	2.87	&	0.61	&	0.07	&	-0.18	&	0.39	&	0.12	&	0.06	\\
					&	fit-bias	&	unif.		&	unif.		&	unif.		& $\delta_{Mc}= 0$		&	"	&	"	&	2.42	&	0.77	&	0.08	&	-0.82	&	0.28	&	0.08	&	0.04	\\
	\hline
	LOCUSS			&	fit-lin		&	unif.		&	unif.		&	Student	& $\delta_{Mc}\neq 0$	&	30	&	0.23	&	4.82	&	0.55	&	0.04	&	-0.44	&	0.26	&	0.18	&	0.04	\\
					&	fit-log	&	log-unif.	&	log-unif.	&	Student	& $\delta_{Mc}\neq 0$	&	"	&	"	&	5.03	&	0.50	&	0.06	&	-0.36	&	0.31	&	0.19	&	0.06	\\
					&	fit-bias	&	unif.		&	unif.		&	unif.		& $\delta_{Mc}= 0$		&	"	&	"	&	4.82	&	0.58	&	0.05	&	-0.67	&	0.23	&	0.16	&	0.06	\\
	\hline
	LOCUSS	(regular)	&	fit-lin		&	unif.		&	unif.		&	Student	& $\delta_{Mc}\neq 0$	&	19	&	0.25	&	4.82	&	0.54	&	0.07	&	-0.24	&	0.36	&	0.17	&	0.06	\\
	\hline
	\end{tabular}
}
\end{table*}

\begin{figure*}
\begin{tabular}{cc}
\includegraphics[width=8.5cm]{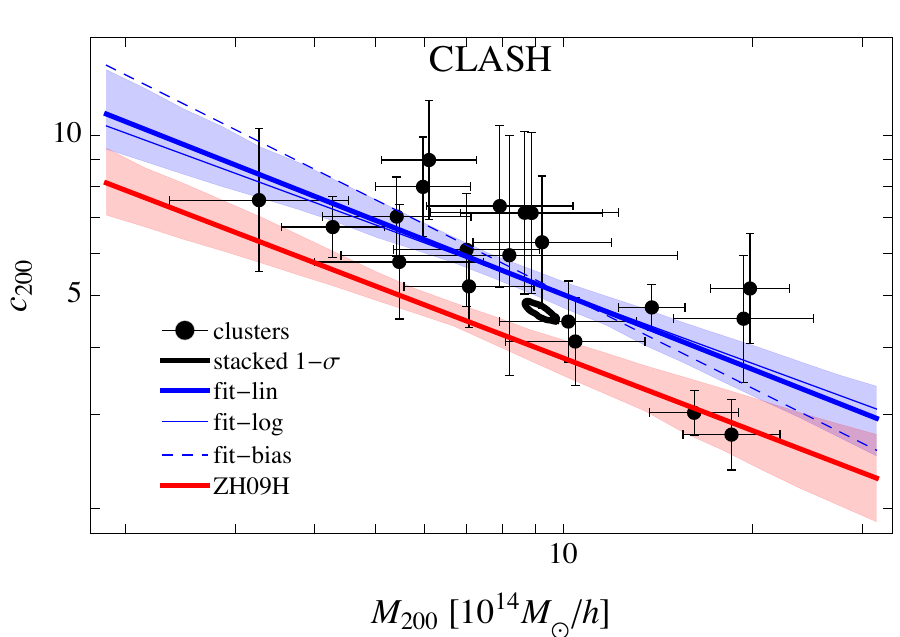} &\includegraphics[width=8.5cm]{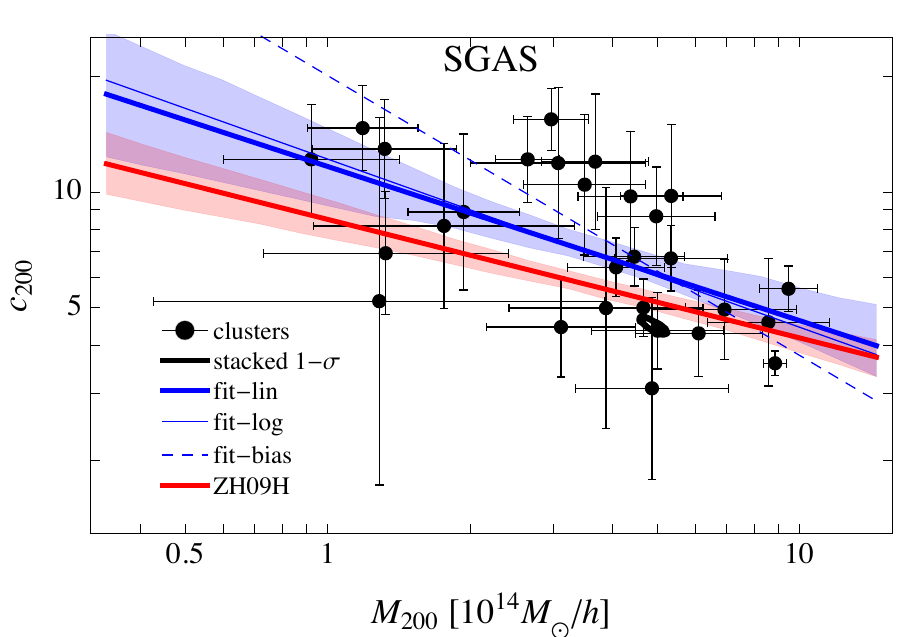}\\
\noalign{\smallskip}  
\includegraphics[width=8.5cm]{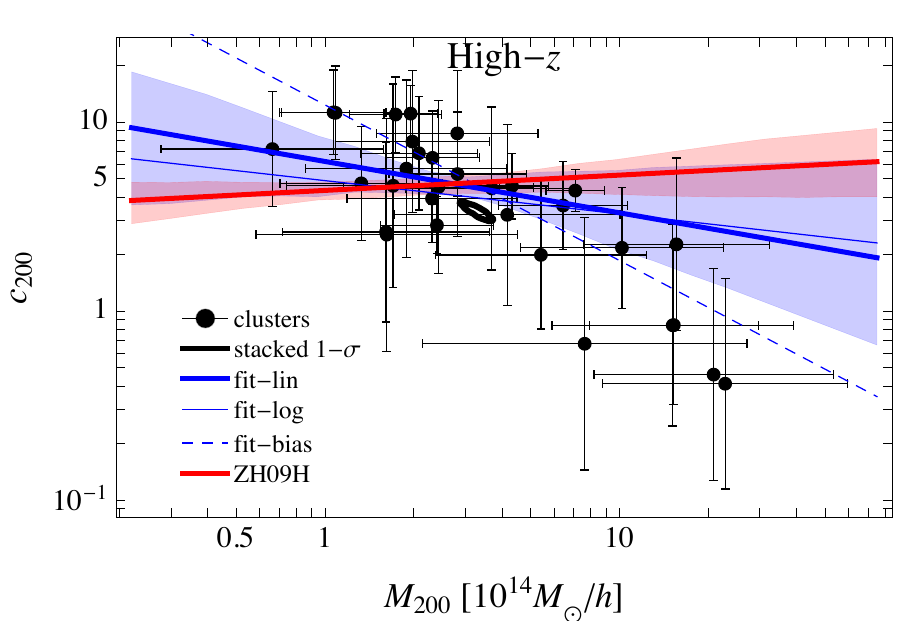} & \includegraphics[width=8.6cm]{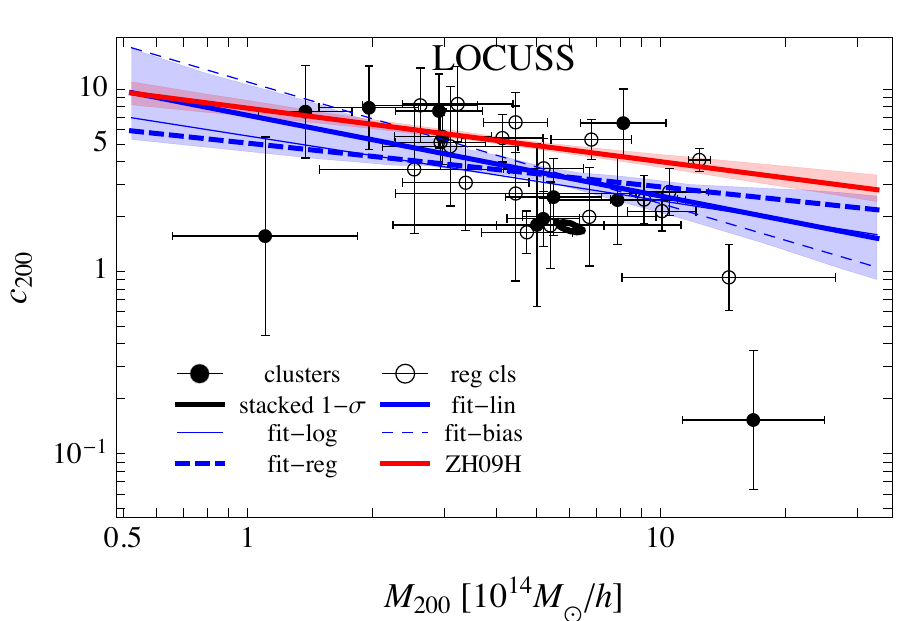}
\end{tabular}
\caption{Concentration, $c_{200}$, versus mass, $M_{200}$. The black points mark the data, the blue graphics represent the fit to the data, and the red graphics represent the theoretical predictions for the observed samples. The thick black contour represents the result from the fit to the stacked profiles. It encloses the 1-$\sigma$ confidence regions for two parameters. The thick, thin, and dashed blue lines correspond to the `fit-lin', `fit-log', and `fit-bias' regression procedure, respectively. The blue area encloses the 1-$\sigma$ region of the `fit-lin' regression. The thick red line corresponds to theoretical prediction based on the $c$-$M$ by \citet{zha+al09} and a Hernquist profile for the BCG galaxy. The top left, top right, bottom left, and bottom right panels illustrate the relation for the CLASH, the SGAS, the high-$z$, and the LOCUSS samples, respectively. For the LOCUSS panel (bottom right), empty circles denote clusters in the regular sub-sample and the thick-dashed line corresponds to the `fit-lin' regression of the regular sub-sample.}
\label{fig_M200_vs_c200_fit}
\end{figure*}

We modelled the observed $c$-$M$ relation with a power law,
\beq
\label{eq_cM_1}
c_{200} =10^\alpha \left( \frac{M_{200}}{M_\mathrm{pivot}}\right)^\beta,
\eeq
as adequate in small redshift and mass intervals. The intrinsic scatter $\sigma$ in the concentration around $c_{200}(M_{200})$ is taken to be lognormal \citep{duf+al08,bha+al13}. 

We studied the observed relation between lensing mass and concentration. This differs from the analysis of the relation between intrinsic mass and concentration since weak lensing estimates are scattered proxies of the true values \citep{ras+al12,se+et14_comalit_I}.

We performed a linear regression in decimal logarithmic ($\log$) variables. If errors are correlated, the observed covariance between the covariate and the response is biased \citep{ak+be96}. Bayesian methods easily account for correlated measurement errors and can provide unbiased estimates of the slope \citep{kel07}.

The intrinsic distribution of $\log M_{200}$ was modelled with a Gaussian function of mean $\mu$ and standard deviation $\tau$. This is a good approximation for flux selected samples of massive clusters \citep{an+be12,ser+al14_comalit_II} and, in general, for regular unimodal distribution \citep{kel07}. 

We chose priors as less informative as possible. We adopted uniform priors for the intercept $\alpha$, and the mean $\mu$. For $\sigma^2$ and $\tau^2$, we considered an inverse Gamma distribution \citep{an+hu10}. We refer to \citet{se+et14_comalit_I} and \citet{ser+al14_comalit_II} for a more extended discussion on assumptions and priors.

To stress the role of the priors and of the uncertainty covariance matrix, we performed the regression examining three cases. 
\begin{description}
\item `fit-lin': as a reference regression, we considered masses and concentrations derived under uniform priors in $M_{200}$ and $c_{200}$. For the slope we assumed a uniform prior on the direction angle $\arctan \beta$, which is equivalent to a Student's $t$ distribution for $\beta$. 
\item `fit-log': as a second case, we considered masses and concentrations determined under priors uniform in logarithmically spaced intervals. As for the reference case, we adopted a Student's $t$ prior on $\beta$ and the full uncertainty covariance matrix.
\item `fit-bias': alternatively, we considered more usual assumptions. In this commonly employed approach, the correlation between measurement errors is neglected and the prior on the slope (instead of the direction angle) is assumed to be uniform. This prior on the slope biases the estimate of $\beta$ high \citep{ser+al14_comalit_II}. At the same time, neglecting the negative correlation between measured quantities makes the measured $c$-$M$ relation steeper. 
\end{description}

The linear Bayesian regression was performed through JAGS\footnote{The program JAGS (Just Another Gibbs Sampler) is publicly available at \url{http://mcmc-jags.sourceforge.net/}.}. Previous application of Bayesian modelling in astronomical contexts can be found in \citet{an+hu12}, \citet{se+et14_comalit_I}, \citet{ser+al14_comalit_II} and references therein. Results are listed in Table~\ref{tab_par_obs} and plotted in Figure~\ref{fig_M200_vs_c200_fit}. 

If we neglect the correlation between measurements (`fit-bias'), results are in agreement with previous analyses, notwithstanding the remaining minor differences in the fitting procedure. Results based on such approaches are flawed. The slope $\beta$ is biased toward the main degeneracy between the measured parameters. The effect is significant for the samples we analyzed. Neglecting the correlation between mass and concentration, estimated slopes are steeper by $\Delta \beta=$0.23, 0.33, 0.55 and 0.23 for the CLASH, SGAS, high-$z$ and LOCUSS sample, respectively.

The correlation between mass and concentration affects the estimate of the scatter too. With the commonly used assumptions made in `fit-bias', $\sigma_{\log}$ is under-estimated by 0.03, 0.05, 0.04 and 0.02 for the CLASH, SGAS, high-$z$, and LOCUSS sample, respectively.

\section{Stacked analysis}
\label{sec_stac}

\begin{figure*}
\begin{tabular}{cc}
\includegraphics[width=8.5cm]{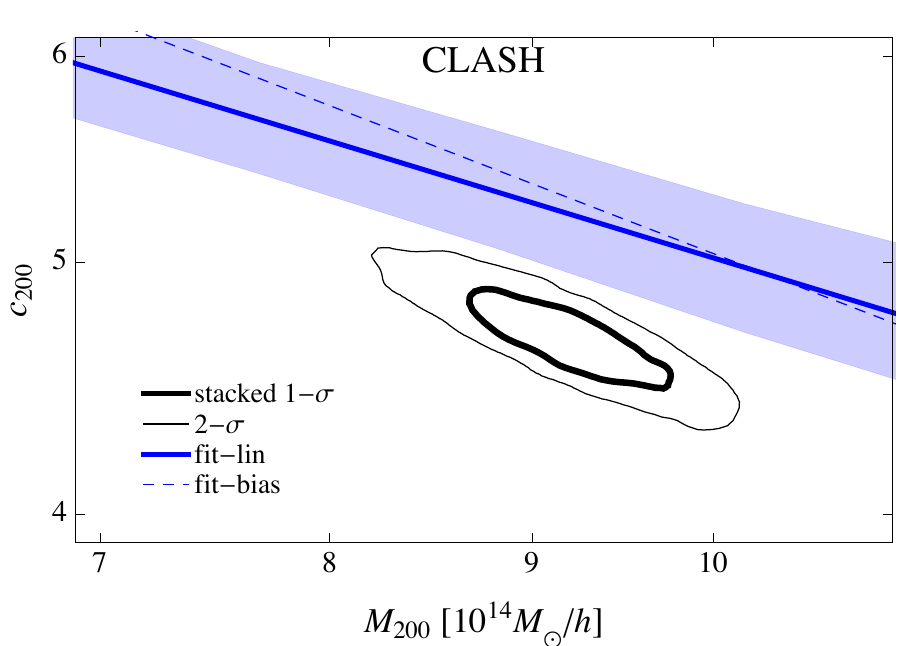} &\includegraphics[width=8.5cm]{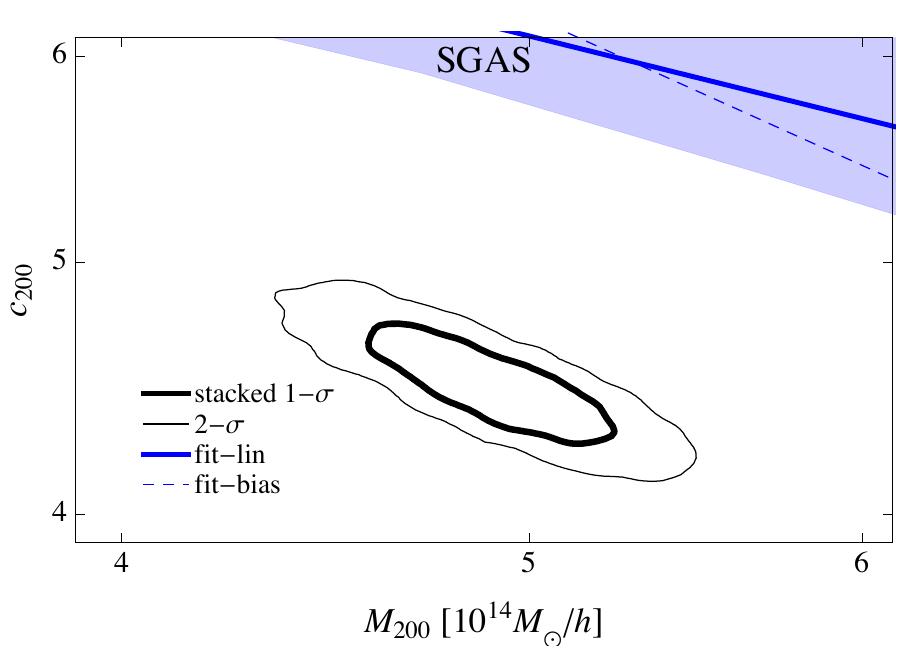}\\
\noalign{\smallskip}  
\includegraphics[width=8.5cm]{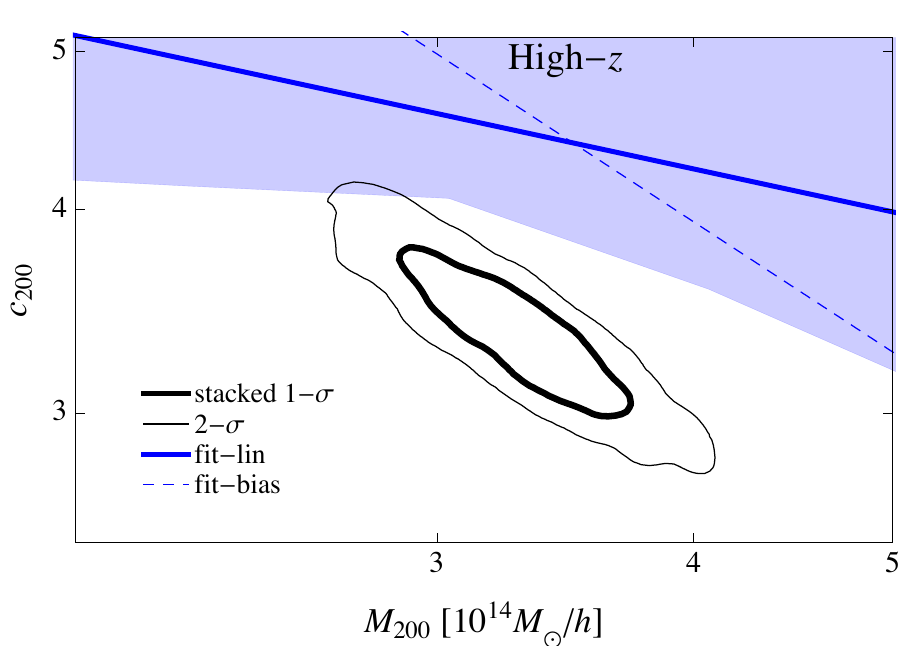} & \includegraphics[width=8.6cm]{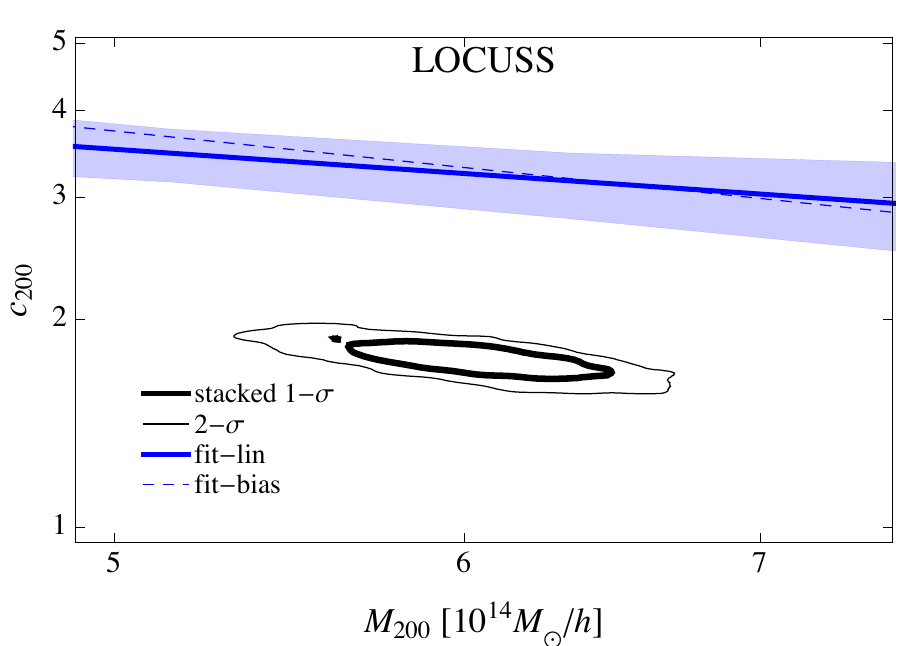}
\end{tabular}
\caption{Concentration, $c_{200}$, and mass, $M_{200}$ of the stacked clusters. The thick and thin black lines enclose the 1- and 2-$\sigma$ confidence regions for two parameters. The blue graphics represent the $c$-$M$ relation fitted to the individual clusters. The thick and dashed blue lines correspond to the `fit-lin', and `fit-bias' regression procedure, respectively. The blue area encloses the 1-$\sigma$ region for the `fit-lin' case. The top left, top right, bottom left, and bottom right panels illustrate the results based on the CLASH, the SGAS, the high-$z$, and the LOCUSS sample, respectively.}
\label{fig_M200_vs_c200_stack}
\end{figure*}

Combining observations from numerous clusters enhances the weak lensing signal. The usual approach consists in stacking the shear measurements \citep{joh+al07,man+al08,ogu+al12,oka+al13,cov+al14} and then fitting a single profile to the stacked signal. An alternative, which we follow here, consists in fitting all shear profiles at once assuming that all clusters have the same mass and concentration \citep{se+co13}. This approach brings some positive features: $i$) we do not have to compute the differential surface density from the shear measurements and the (would be) induced redshift dependence is removed; $ii$) we can fit the reduced shear instead of the shear, which allows us to extend the analysis even in the inner regions; $iii$) this approach avoids the choice between co-adding either in physical or in rescaled radius, which might make the stacked profile shallower; $iv)$ we can still use the strong lensing constraints. The related $\chi^2$ function is
\beq
\chi^2_\mathrm{Stacked}=\sum_j \chi_{\mathrm{GL},j}^{2} (z_j; M_{200}, c_{200}),
\eeq
where the sum extends over the sample, $z_j$ is the redshift of the $j$-th cluster and $\chi^2_\mathrm{GL}$ is defined in Eq.~(\ref{eq_chi_GL}). As for our reference case, $M_{200}$ and $c_{200}$ were determined assuming uniform priors. 

Results are summarised in Table~\ref{tab_stacked}. Our estimates compare well with previous works. \citet{ume+al14} found $M_{200}\sim9.3 \times 10^{14}M_\odot/h$ and $c_{200}\sim 4.0$ from the stacked shear profiles of 20 CLASH clusters.  \citet{ogu+al12} found $M_{200}\sim4.1 \times 10^{14}M_\odot/h$ and $c_{200}\sim 4.9$ for 25 clusters in the SGAS sample. 

Lensing is highly non linear. More massive lenses can be observed with a larger signal to noise ratio and they might have a larger weight in the stacked analysis. The mass of the stacked halo $M_{200}^\mathrm{stacked}$ is then systematically larger than the median mass of the sample, even though still consistent within the quoted uncertainties. 

However, the weight of cluster with higher signal-to-noise ratio depend on the stacking technique. Based on numerical simulations \citep{mcc+al11,bah+al12}, \citet{oka+al13}  found that stacking in physical length units does not bias the measurement of the cluster mass and concentration. In fact, the weight factor is mass-independent for stacking in physical length units, i.e., non-rescaled with the over-density radius \citep{oka+al13,ume+al14}. On the other hand, stacking in rescaled radial units gives biased extraction of mass and concentration parameters due to the mass-dependent weight factor  \citep{oka+al13}.

The mass of the stacked halo determines the scale where the $c$-$M$ relation of the sample is normalised. We can see from Figs.~\ref{fig_M200_vs_c200_fit} and ~\ref{fig_M200_vs_c200_stack} that notwithstanding the different assumptions used to fit the data, variously fitted $c$-$M$ relations intersect at $\sim M_{200}^\mathrm{stacked}$.

The concentration of the stacked profile, $c_{200}^\mathrm{stacked}$, is smaller than the value predicted at that mass by the $c$-$M$ relation fitted to the sample, see Fig.~\ref{fig_M200_vs_c200_stack}. $c_{200}^\mathrm{stacked}$ is likely biased low by our requirement to fit all profiles with a single concentration. As said before, more massive lenses, which are usually less concentrated, have the larger influence in the stacked analysis. Furthermore, a low concentration may be more suited to fit a variegated sample.

The confidence regions of the stacked mass and concentration align with the slope of the biased fit, see Fig.~\ref{fig_M200_vs_c200_stack}. If we do not correct for correlation, the retrieved slope of the $c$-$M$ relation goes along with the main degeneracy between mass and concentration. This bias is exacerbated by the small mass range that usually characterises lensing sample and by the quite shallow intrinsic slope of the $c$-$M$ relation.

\begin{table}
\centering
\caption{Constraints from the stacked analysis on the typical mass and concentration of the samples. Col.~1: sample; col.~2: stacked mass $M_{200}$ in units of $10^{14}M_\odot/h$; col.~3: stacked concentration $c_{200}$; col.~4: correlation factor between $M_{200}$ and $c_{200}$. Quoted central values and uncertainties are bi-weight estimators of the posterior probability distribution.}
\label{tab_stacked}
\begin{tabular}[c]{l r@{$\,\pm\,$}lr@{$\,\pm\,$}lc}
        \hline
        	Sample			& \multicolumn{2}{c}{$M_\mathrm{200}$} & \multicolumn{2}{c}{$c_\mathrm{200}$} & $\delta_{Mc}$  \\
					& \multicolumn{2}{c}{[$10^{14}M_\odot/h$]} & \multicolumn{2}{c}{ } &  \\
        \hline
      	CLASH		&	9.20	& 	0.40	&	4.69	&	0.15	&	-0.89	\\
	SGAS		&	4.92	&	0.23	&	4.49	&	0.16	&	-0.86	\\
	High $z$		&	3.30	&	0.28	&	3.39	&	0.27	&	-0.91	 \\
	LOCUSS		&	6.06	&	0.27	&	1.75	&	0.08	&	-0.74	\\
	\hline
	\end{tabular}
\end{table}

\section{Theoretical predictions}
\label{sec_theo}

\begin{figure*}
\begin{tabular}{cc}
\includegraphics[width=8.5cm]{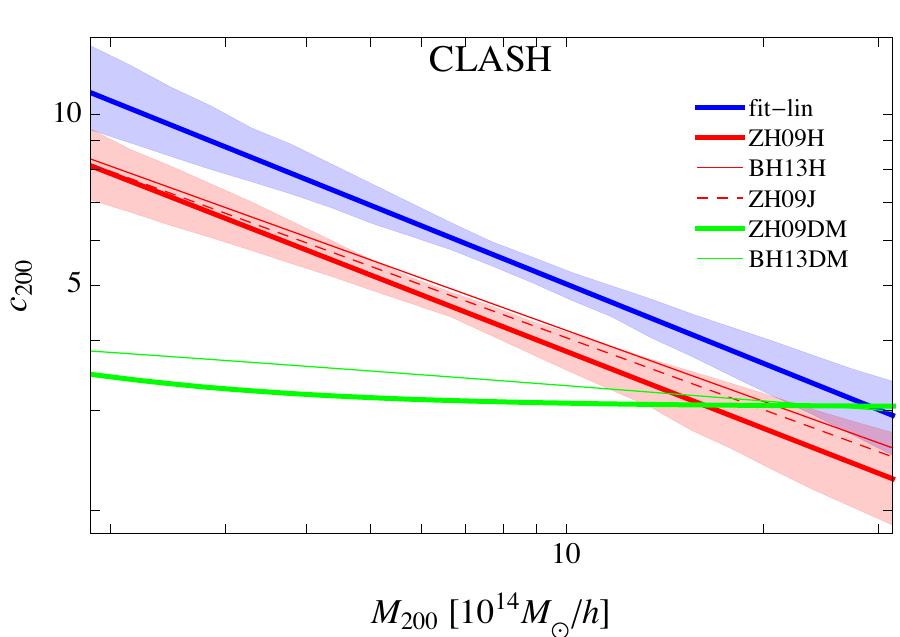} &\includegraphics[width=8.5cm]{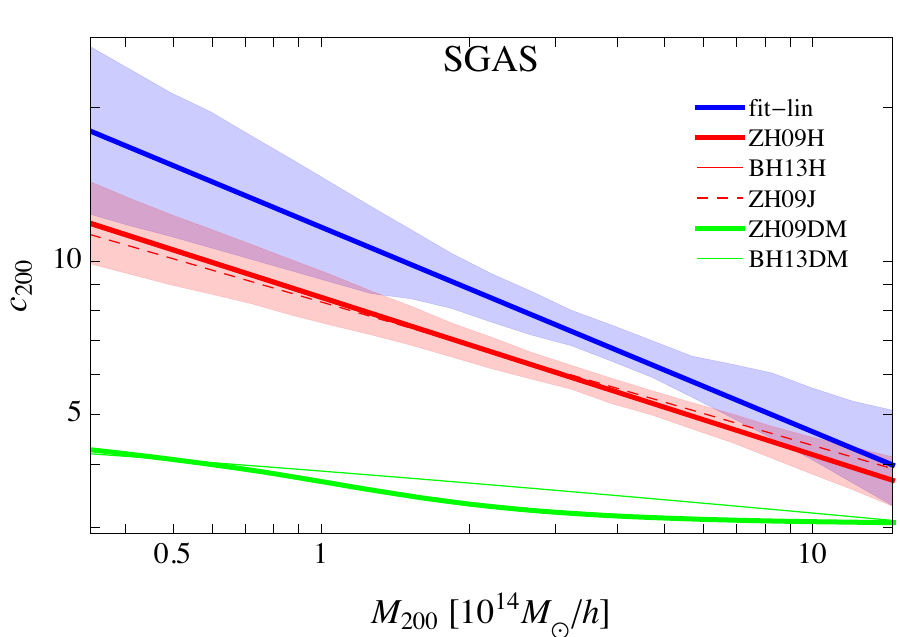}\\
\noalign{\smallskip}  
\includegraphics[width=8.5cm]{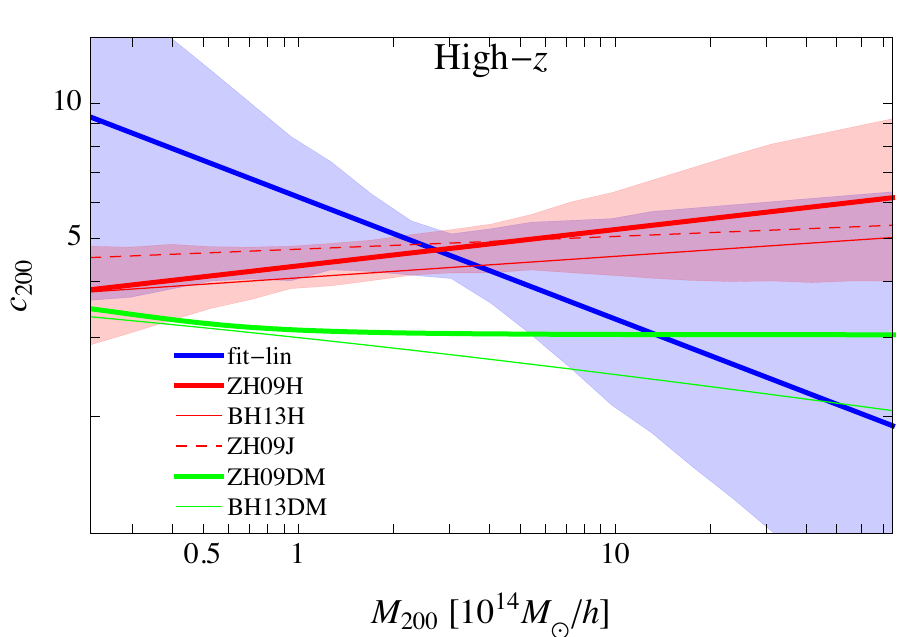} & \includegraphics[width=8.6cm]{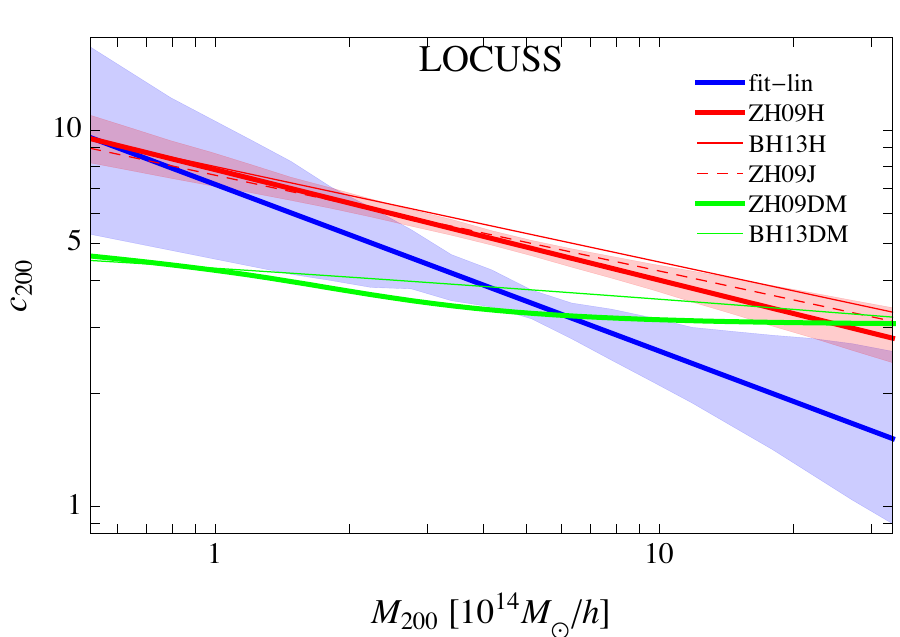}
\end{tabular}
\caption{Theoretical predictions for concentration, $c_{200}$, versus mass, $M_{200}$. The red graphics represent the theoretical predictions for the observed samples, the green lines the intrinsic $c$-$M$ relations for dark matter haloes, and the blue line represents the `fit-lin' regression to the real data. The blue area encloses the 1-$\sigma$ region for the `fit-lin' case. The thick, thin, and dashed red lines correspond to ZH09H, BH13H, and ZH09J (see Sec.~\ref{sec_theo} for details), respectively. The thick and thin green lines correspond to the input $c$-$M$ relations of ZH09DM and BH13DM at the median redshift of the sample, respectively. The top left, top right, bottom left, and bottom right panels illustrate the results for the CLASH, the SGAS, the high-$z$, and the LOCUSS sample, respectively.}
\label{fig_M200_vs_c200_th}
\end{figure*}

\begin{table}
\caption{Predicted $c$-$M$ relation for CLASH-, SGAS-, high-$z$-, and LOCUSS-like samples. Col.~1: sample; col.~2: input $c$-$M$ relation for the dark-matter haloes. `ZH09' stands for \citet{zha+al09}; `BH13' stands for \citet{bha+al13}. Col.~3: assumed density profile of the BCG (`H' and `J' stand for either a Hernquist or a Jaffe profile, respectively). Cols.~4, 5, and 6: normalisation, slope, and intrinsic scatter of the $c$-$M$ relation, modelled as a power law. The pivot masses are as in Table~\ref{tab_par_obs}. For an easier comparison, we report again the results of the reference fitting procedure to the real samples (`Observed'), i.e., the `fit-lin' case of Table~\ref{tab_par_obs}. Quoted values are bi-weight estimators of the posterior probability distribution.}
\label{tab_par_th}
\resizebox{\hsize}{!} {
\begin{tabular}[c]{l  l c r@{$\,\pm\,$}lr@{$\,\pm\,$}lr@{$\,\pm\,$}l}
        \hline
	Sample			& $c$-$M$	& BCG	& \multicolumn{2}{c}{$\alpha$} & \multicolumn{2}{c}{$\beta$} & \multicolumn{2}{c}{$\sigma$}  \\
        \hline
	CLASH		&	ZH09	&	H		&	0.61	&	0.03		&	-0.45	&	0.12	&	0.11	&	0.02	\\
				&	ZH09 	&	J		&	0.63	&	0.02		&	-0.41	&	0.10	&	0.10	&	0.02	\\
				&	BH13	&	H		&	0.64	&	0.02		&	-0.40	&	0.10	&	0.12	&	0.02	\\
				& \multicolumn{2}{l}{Observed} &	0.73	&	0.02	&	-0.46	&	0.11	&	0.08	&	0.02	\\
	\hline
	SGAS		&	ZH09	&	H		&	0.74	&	0.02		&	-0.31	&	0.07	&	0.15	&	0.02	\\
				&	ZH09 	&	J		&	0.75	&	0.02		&	-0.28	&	0.08	&	0.14	&	0.02	\\
				&	BH13	&	H		&	0.74	&	0.02		&	-0.30	&	0.07	&	0.14	&	0.02	\\
				& \multicolumn{2}{l}{Observed}&	0.82	&	0.03	&	-0.40	&	0.14	&	0.12	&	0.02	\\
	\hline
	high-$z$		&	ZH09	&	H		&	0.66	&	0.04		&	0.08		&	0.11	&	0.26	&	0.05	\\
				&	ZH09 	&	J		&	0.68	&	0.04		&	0.03		&	0.11	&	0.24	&	0.05	\\
				&	BH13	&	H		&	0.63	&	0.04		&	0.06		&	0.10	&	0.27	&	0.05	\\
				& \multicolumn{2}{l}{Observed}&	 0.69	&	0.06	&	-0.27	&	0.35	&	0.12	&	0.05	\\
	\hline
	LOCUSS		&	ZH09	&	H		&	0.70	&	0.02		&	-0.30	&	0.07	&	0.10	&	0.01	\\
				&	ZH09 	&	J		&	0.71	&	0.02		&	-0.25	&	0.07	&	0.11	&	0.01	\\
				&	BH13	&	H		&	0.73	&	0.02		&	-0.25	&	0.07	&	0.10	&	0.01	\\
				& \multicolumn{2}{l}{Observed}&	 0.55	&	0.04	&	-0.44	&	0.26	&	0.18	&	0.04	\\
	\hline

	\end{tabular}
}
\end{table}

We do not observe the intrinsic $c$-$M$ relation. Firstly, we measure the lensing mass and concentration from projected maps. These are scattered and even slightly biased proxies of the intrinsic 3D mass and concentration \citep{men+al10,gio+al12b,se+et14_comalit_I}. The main sources of bias and scatter are the halo triaxiality \citep{ji+su02,des+al14} and the presence of substructures within the host halo virial radius \citep{gio+al12b}.

Secondly, the observed $c$-$M$ relation has also to differ from the outputs of numerical simulations of dark matter haloes. Baryons cool and sink into the inner regions and gravitationally contract the dark matter distribution. This steepens the $c$-$M$ relations since the effect is more prominent at low masses \citep{fed12,gio+al12b}. Furthermore, the presence of a BCG strongly affects the very inner slope of the cluster \citep{gio+al14}.

Finally, selection effects severely impact the measured mass--concentration relation \citep{ogu+al12,men+al14}. Strong lensing clusters may have a concentration $\sim 20-30$ per cent higher than the average, at fixed mass \citep{gio+al14}.  

The above effects can be taken into account with a semi-analytical approach \citep{gio+al12a}. In order to describe the expected $c$-$M$ relation of the different observed cluster samples, we performed various simulations with the MOKA code. In the following, we briefly summarise some relevant aspects whereas we refer to \citet{gio+al12a,gio+al14} for a detailed description of the implementation of the halo triaxiality, the substructures distribution, the presence of the BCG, and the baryonic physics. 

To create a cosmological cluster sample, we randomly draw halo masses from the \citet{sh+to99} mass function saving all haloes above $10^{14} M_{\odot}/h$. For six different redshifts (0.19, 0.29, 0.35, 0.45, 0.55, and 0.90), the number of haloes is created to match the count of collapsed objects present on the whole sky between $z-\Delta z/2$ and $z+\Delta z/2$ (with $\Delta z=0.01$). To increase the statistical sample, for each redshift we performed eight different realisations. The final number of clusters for each redshift bin is 15053, 27222, 34009, 41853, 45510, and 36499, respectively. 

We considered two input $c$-$M$ relations for the dark-matter halos: $i$) `ZH09DM', in which the concentrations follow the empirical model developed in \citet[ ZH09]{zha+al09} and based on the mass accretion histories of dark matter halos; $i$) `BH13DM', in which the mass and concentration are related through the relation found in \citet[ BH13]{bha+al13} and based on gravity-only simulations.

Once the cluster mass samples have been created, we run the MOKA code for each redshift with different input parameters, creating three simulated samples:
\begin{description}
\item `ZH09H', in which the input $c$-$M$ follows  `ZH09DM', and the BCG follows an \citet[ H]{her90} profile; 
\item `ZH09J',  where the input $c$-$M$ follows  `ZH09DM', and the BCG follows a \citet[ J]{jaf83} profile; 
\item `BH13H',  where the $c$-$M$ goes like `BH13DM' and the BCG is modelled with an Hernquist profile. 
\end{description}

As done in \citet{gio+al14}, during each run we computed the convergence, the reduced tangential shear, and the potential maps of the whole triaxial and substructured cluster. We then randomly located in the field of view, that goes up to the cluster virial radius, a sample of background galaxies with a density of 30 galaxies per square arcminute and extracted the reduced tangential shear profile. 2D lensing mass and concentration were finally computed through a best fit procedure based on the NFW functional, see Eq.~(\ref{eq_chi_WL}). The high density of background sources was chosen because we are more interested in studying how selection and other physical effects affects the $c$-$M$ relation rather than estimating the accuracy which the projected relation can be recovered within.

For each MOKA cluster we also extracted the information about the size of the Einstein radius, the shape of the projected mass density, and the potential ellipticity within $R_{500}$ that we could compare with the X-ray morphology of the observed cluster when available.   

For each observed cluster sample (CLASH, SGAS, high-$z$, and LOCUSS) we created a solid MOKA sample selecting for each observed cluster all MOKA haloes at the nearest redshift with an estimated 2D mass consistent -- at least within the error bars -- with the observed one. In selecting the MOKA clusters, in order to be as consistent as possible with the real selection function, we included additional selection criteria for each observed cluster when more information was available, like: the size of the Einstein radius, which is the case of the CLASH \citep{mer+al14}, the SGAS \citep{ogu+al12}, and some high-$z$ clusters \citep{se+co13}; the X-ray morphology for the CLASH sample \citep{pos+al12}; the optical shape for the SGAS sample \citep{ogu+al12}. For the LOCUSS sample, at the light of the work presented by \citet{ric+al10}, we selected only MOKA clusters with an effective Einstein radius of at least 5\arcsec (for sources at $z_\mathrm{s}=2$).

Based on the above criteria, we extracted 1024 CLASH-, SGAS-, high-$z$-, and LOCUSS-like samples. Each sample was fitted with the unweighted BCES($X_2|X_1$) estimator \citep{ak+be96}. Since we performed unweighted regressions, the result is not particularly dependent on the adopted regression scheme. The parameters of the $c$-$M$ relations were finally estimated as the bi-weight estimators of the distributions of best fitting parameters. The theoretical predictions for the considered samples are summarised in Table~\ref{tab_par_th} and plotted in Fig.~\ref{fig_M200_vs_c200_th}. We remark that the prediction for the CLASH sample cannot be directly compared with \citet{men+al14}, who also considered the redshift dependence of the $c$-$M$ relation. Constraints on mass and redshift evolution can be highly degenerate.

The interplay between baryons and dark matter, the presence of the BCG, and mostly selection effects concur to significantly increase the observed concentrations, mostly at smaller masses. As a result the observed 2D $c$-$M$ relations are much steeper than the corresponding 3D relations for dark matter only halos, see Fig.~\ref{fig_M200_vs_c200_th}. Adiabatic contraction increases the density concentration in the inner regions, mostly at lower masses where the effect of baryons is larger \citep{fed12}. Selection effects preferentially include over-concentrated clusters. The concentration of strong lens clusters may be larger by $\sim 20-30$ per cent than the average, at fixed mass \citep{gio+al14}. Clusters exhibiting a nearly circular morphology in the plane of the sky are likely triaxial objects elongated along the line of sight, for which the measured 2D concentration and mass are significantly larger than the intrinsic 3D values \citep{hen+al07,og+bl09}. 

The shape of the BCG plays a role too. A BCG modelled as a Jaffe profile steepens the $c$-$M$ relation more than a Hernquist model. If the $c$-$M$ is modelled after \citet{bha+al13}, the expected slope is slightly flatter than in the case based on \citet{zha+al09}. These variations are within the statistical uncertainties.

We found that predicted theoretical slopes are in very good agreement with observations, see Table~\ref{tab_par_th}. The CLASH and the SGAS samples are slightly over-concentrated but still consistent within errors with predictions. Sources of possible disagreement are discussed in the next section.

Apart from the LOCUSS sample, observed scatters are smaller than the predicted values, but as for the normalisation, the difference is not statistically significant.

\section{Discussion}
\label{sec_disc}

In this section, we review some additional aspects that impact the $c$-$M$ relation. We first discuss some problematics concerning the process of measurement and then some theoretical aspects.

\subsection{Multiple matter components}

The concentration relates a global property of the cluster, $r_{200}$, to the local slope in the inner region. A concentration can be attributed to any form of density profile by defining it as the ratio of the outer `virial' radius and the radius at which the logarithmic slope is -2. 

The concept itself of concentration might be then ill-defined in irregular clusters. In fact, the measurements and the properties of the concentration are strongly dependent on the assumed halo shape for complex morphologies \citep{pra+al11,du+ma14,men+al14}. 

A further complication is provided by the differentiated nature of the cluster components. The baryons are distributed among intracluster stars, galaxies, a possible dominant BCG, which contribute to the total matter density mainly in the inner regions, and diffused hot gas. In relaxed clusters, the gas follows the gravitational potential and its distribution has a rounder shape and a flatter slope than the dark matter. In merging clusters, the gas may be displaced from the baryons and the dark matter, which makes the definition of concentration even trickier.

Whereas in numerical simulations and semi-analytical studies it is more immediate to define a global halo and to study its properties, the complex nature of the real clusters may be better addressed differentiating it in multi-components. A possible solution can come from lensing analyses that single out the gravitational action of the gas, the galaxies, and the dark matter \citep{ser+al10b}. This requires multi-wavelength observations from the X-ray to the optical to the radio and observations in both the strong and the weak lensing regime \citep{ser+al13}. 

X-ray studies based on the assumption of hydrostatic equilibrium (HE) can also measure the concentration of either the total or the dark-matter only profiles \citep{ett+al10}. However, the bias in the mass measurement associated to the HE hypothesis is significant \citep{ras+al12,se+et14_comalit_I}. Masses are usually biased low by $\gs 20$ per cent whereas the different level of deviation from equilibrium of different regions (outer regions are supposed to be less relaxed) may affect the estimate of the concentration.

\subsection{Observational uncertainties}

The measurement of mass and concentration is challenging even for regular clusters. The level of known systematics in weak lensing mass determinations is of the order of $\sim 10$ per cent \citep{wtg_III_14,ume+al14} but differences in the mass reported by independent groups are as large as 40 per cent \citep{se+et14_comalit_I}.

Detailed comparisons show that weak lensing masses derived with independent methods may be at the same time well correlated but considerably off-set in absolute values \citep{wtg_III_14}. 

Key sources of systematic error in cluster masses are due to: robustness of the shape measurements; uncertainties in the source redshift distribution; dilution/contamination effects
by either foreground or cluster member galaxies; assumptions in the fitting procedure (different radial ranges, different fixed concentrations used, different choices for the centre location). \citet{oka+al13} critically revised a previous analysis to estimate contamination effects and systematics in shape measurements. Effects on mass and concentration were found to be as large as 40 per cent in single clusters.

We relied on published shear profiles and previously estimated source redshifts and we could not assess the impact of the previous effects on the estimated mass-concentration relations.

\subsection{Weak vs. strong lensing}

Accurate estimates of the concentration require the study of the weak lensing signal in the outer regions together with a detailed analysis of the strong lensing systems in the inner regions. The constraints from the two regimes should be properly weighted. For the strong lensing regime, we considered only the position of the effective Einstein radius. This constraint does not embody all the information from the multiple image systems and might down-weight the strong lensing contribution. On the other hand, the position of the strong lensing systems is contingent on the very local matter distribution which can bias the measurement of the concentration.

As an example, let us consider the CLASH sample.  Based on WL only analyses of the clusters, we obtained concentrations larger by $9\pm 50$ per cent than the estimates including the additional SL constraint. The very large scatter shows that the contribution of the strong lensing information is crucial in the estimate of the concentration. 

This is further stressed by the comparison of our SL+WL estimates to the results of \citet{mer+al14}.  \citet{mer+al14} gave full weight to the strong lensing regime by inferring the position of the critical lines at the various source redshifts from the position of the multiple images. The extent of the SL constraint shows up in the accuracy. Our typical statistical uncertainty on $c_{200}$ is of order of $\sim 40$ per cent, whereas errors in \citet{mer+al14} are smaller, $\sim 30$ per cent. 

The determinations of $c_{200}$ in \citet{mer+al14} are smaller by $40 \pm 30$ per cent. The scatter is significant and the typical deviation is of the order of our statistical error. Nevertheless, concentrations from \citet{mer+al14} are systematically smaller, which can reduce the tension we found for the CLASH sample with respect to theoretical predictions.

\subsection{The shape of the $c$-$M$}

The theoretical $c$-$M$ relation is still debated even in the context of dark-matter only models. Results from numerical simulations depend on the mass resolution, on the simulation volume \citep{pra+al11}, and on the binning and fitting procedures \citep{me+ra13}. At low masses and redshifts, the $c$-$M$ relation of dark matter simulated haloes is well fitted by a power-law \citep{net+al07,mac+al08,gao+al08,duf+al08}. There are some indications of a flattening and upturn of the relation with increasing mass and redshifts \citep{pra+al11} but the presence and the extent of such feature is still questioned \citep{me+ra13}.

On theoretical grounds and as probed by numerical simulations, the concentration is related to the halo assembly history \citep{nav+al97,bul+al01,zha+al09,gio+al12c}. Since the initial density peak and the halo mass accretion history are strictly connected, concentrations in $\Lambda$CDM and self-similar cosmologies can then be accurately described by a universal function of the peak height $\nu$ \citep{bha+al13,lud+al14}. 

The theoretical understanding of the shape of the  $c$--$\nu$ relation is crucial. A simple power-law might not be a good approximation over a wide range of $\nu$ and an upturn might show up at high $\nu$ \citep{di+kr15}.

\subsection{Cosmological parameters}

As a consequence of the link with the halo assembly history, the $c$-$M$ relation strongly depends on the cosmological framework. It has also been proposed to constrain cosmological parameters \citep{buo+al07,ett+al10,gio+al12b}. The normalisation of the power spectrum $\sigma_8$ and the dark matter content strongly affect the overall amplitude of the relation \citep{du+ma14}. A residual dependence of concentration on the local slope of the matter power spectrum might affect both the amplitude and the shape of the $c$--$\nu$ relation \citep{di+kr15}. The dark energy equation of state parameter has an effect too, mainly for lower mass haloes  \citep{kwa+al13}.

Whatever the shape of the $c$--$\nu$ functional form is, the accurate knowledge of cosmological parameters is crucial to predict the relation between masses and concentrations.

\subsection{Baryonic physics}

Baryonic physics contributes to shape the cluster density profile in the inner regions through the competing effects of cooling and feedback processes \citep{gne+al04,roz+al08,duf+al10,deb+al13}. The cooling of baryons and the consequent shrinking effect experienced by dark matter make haloes more concentrated in the inner regions. The mass--concentration relation after this adiabatic contraction is steeper than the theoretical expectation for dark-matter only haloes, because star formation is fractionally more efficient in low-mass objects \citep{mos+al10}. However, the effect is non-vanishing at all masses, which determines a larger normalisation \citep{fed12}. 

On the other hand, other baryonic processes mitigate the effects of contraction. The baryon fraction in the inner regions of clusters is lowered by AGN (active galactic nucleus) feedback, or extremely efficient feedback from massive stars \citep{kil+al12}. This determines shallower inner density profiles and lower concentrations \citep{duf+al10,mea+al10}. The counterbalancing actions of cooling and feedback are still debated. They have to be included in a consistent picture that accounts at the same time for a steep $c$-$M$ relation and for the observed stellar fraction in galaxy clusters \citep{duf+al08}. 




\section{Conclusions}
\label{sec_conc}

The relation between mass and concentration in galaxy clusters summarises important features of their formation and evolution history. In the $\Lambda$CDM model of structure formation, observations of lensing clusters were considered in tension with theoretical predictions. Massive clusters appeared to be over-concentrated and the $c$-$M$ relation much steeper than predicted. We discussed critically some major sources of disagreement.

The measurements of mass and concentration are strongly (anti-)correlated. If the errors are correlated, the slope in the scaling relation is biased \citep{ak+be96}. This is a very important effect in the linear regression of mass vs. concentration. When we correct for it, the observed relation is significantly flatter. The effect is sample-dependent but it is always sizeable, of the order of $\Delta \beta \sim$0.2--0.3 for well observed samples (CLASH, SGAS or LOCUSS) and even larger for less constrained samples ($\Delta \beta \sim 0.5$ for the high-$z$ sample).  

On the theoretical side, the $c$-$M$ we can measure from lensing observations is unlike the scaling relation between masses and concentrations of dark matter halos \citep{gio+al12b,gio+al14,men+al14}. Firstly, due to projection effects, quantities measured with lensing are scattered proxies of the true quantities \citep{ras+al12,se+et14_comalit_I}. Secondly, baryonic physics impacts the relation. Thirdly, selection effects may significantly steepen the relation. The latter effect is the most important in today samples of lensing clusters, which are usually selected in X-ray flux or based on their strong lensing properties. As a result, the theoretical prediction for the $c$-$M$ is significantly steeper than the corresponding input relation of dark matter haloes.

If all of the previous effects are accounted for, the tension between observations and predictions is mostly solved. The determination of the mass--concentration relation can be biased by centre offset, dilution/contamination effects, uncertain source redshift distributions, and shape noise. We relied on published shear profiles, which may be affected by these systematics. However, we found consistent results among different samples of clusters analysed with independent methods. This stresses the importance of statistical biases and selection effects in the determination of the $c$-$M$.

A full understanding of systematics enables us to use the $c$-$M$ relation to study some subtler effects, such as the dependency of the relation on the cosmological parameters or the role of AGN feedback in the cluster physics.

The present paper is in line with recent work by the CLASH team \citep{mer+al14,men+al14}. \citet{men+al14} estimated the theoretical expectation for the $c$-$M$ relation in a CLASH-like sample. They derived lensing-like concentrations and masses from numerical simulations after accounting for the CLASH selection function based on X-ray morphology. The simulated sample could reproduce the observational properties of the CLASH clusters, in particular their ability to produce strong lensing effects and their X-ray regularity. The simulations were then analysed in 2D to account for possible biases in the lensing reconstructions due to projection effects. The theoretical $c$-$M$ relation and the $c$-$M$ relation derived directly from the CLASH data were found to be in excellent agreement \citep{mer+al14}.

We limited our analysis to lensing samples. Nevertheless, selection effects and statistical biases play the same role also for the $c$-$M$ relations determined with X-ray clusters, which are affected by similar problematics \citep{co+na07,fed12}.

To highlight the effects of selection and statistical biases, we did not consider the redshift evolution of the $c$-$M$ relation. Due to selection limits, clusters at larger redshifts may be more massive. This can mimic a mass dependent effect. However, the study of the evolution requires a very accurate knowledge of the cluster mass function and of the selection function \citep{an+co14}. Ignoring them can led to large biases in the derived evolution. 

Problems connected to redshift evolution can be skipped with the approach we took in the present paper. We did not try to model the dependence on $z$ of the observed concentrations, but we focused on the determination of the $c$-$M$ relation for an observed sample characterised by a given redshift distribution. For a given sample and its specific selection function, the effect of the redshift distribution of the members is already entwined in the slope and normalisation of the $c$-$M$ relation. We then determined the $c$-$M$ relation of the sample rather than the universal $c$-$M$ relation, which depends on $z$. We consistently compared measurements to theoretical predictions tuned to the properties of the observed samples, among which the redshift distribution.

\section*{Acknowledgements}
MS thanks Keiichi Umetsu for useful discussions and Adi Zitrin and Julian Merten for providing useful information on the strong lensing analysis of the CLASH clusters. LM and MS acknowledge financial contributions from contracts ASI/INAF I/023/12/0, by the PRIN MIUR 2010-2011 `The dark Universe and the cosmic evolution of baryons: from current surveys to Euclid' and by the PRIN INAF 2012 `The Universe in the box: multiscale simulations of cosmic structure'. CG's research is part  of the project GLENCO, funded under the European Seventh Framework Programme, Ideas, Grant Agreement n.~259349.


\setlength{\bibhang}{2.0em}

\end{document}